\documentclass[twocolumn]{aastex631}

\usepackage{derivative}
\usepackage{amsmath}

\accepted{July 12, 2022}
\submitjournal{ApJ}

\shorttitle{Planetesimal Dynamics}
\shortauthors{Guo \& Kokubo}
\graphicspath{{./}{figures/}}

\begin{document}

\title{Planetesimal Dynamics in the Presence of a Giant Planet II:\\
Dependence on Planet Mass and Eccentricity}

\correspondingauthor{Kangrou Guo} 

\email{carol.kwok@grad.nao.ac.jp}

\author[0000-0001-6870-3114]{Kangrou Guo}
\affiliation{The University of Tokyo, 113-8654, 7 Chome-3-1 Hongo \\
Bunkyo City, Tokyo, Japan}

\affiliation{National Observatory of Japan, 181-8588, 2 Chome-21-2 Osawa \\
Mitaka City, Tokyo, Japan}

\author[0000-0002-5486-7828]{Eiichiro Kokubo}
\affiliation{The University of Tokyo, 113-8654, 7 Chome-3-1 Hongo \\
Bunkyo City, Tokyo, Japan}

\affiliation{National Observatory of Japan, 181-8588, 2 Chome-21-2 Osawa \\
Mitaka City, Tokyo, Japan}

\begin{abstract}
The presence of an early-formed giant planet in the protoplanetary disk has mixed influence on the growth of other planetary embryos. 
Gravitational perturbation from the planet can increase the relative velocities of planetesimals at the mean motion resonances to very high values and impede accretion at those locations. 
However, gas drag can also align the orbital pericenters of equal-size planetesimals in certain disk locations and make them dynamically quiet and "accretion-friendly" locations for planetesimals of similar sizes. 
Following the previous paper, where we investigated the effect of a Jupiter-like planet on an external planetesimal disk, we generalize our findings to extrasolar planetary systems by varying the planet parameters. 
In particular, we focus on the dependence of the planetesimal relative velocities on the mass and eccentricity of the existing planet. 
We found that the velocity dispersion of identical-mass particles increases monotonically with increasing planet mass. 
Meanwhile, the dependence of the relative velocity between different-mass planetesimals on their mass ratio becomes weaker as the planet mass increases. 
While the relative velocities generally increases with increasing planet eccentricity, the velocity dispersion of smaller-mass particles ($m \lesssim 10^{18}~\rm{g}$) is almost independent of planet eccentricity owing to their strong coupling to gas. 
We find that the erosion limits are met for a wider range of parameters (planet mass/eccentricity, planetesimal mass ratio) when the planetesimal size decreases.
Our results could provide some clues for the formation of Saturn's core as well as the architecture of some exoplanetary systems with multiple cold giant planets.

\end{abstract}



\section{Introduction} \label{sec:intro}

The accretion of planetesimals during the planet-forming stage in a protoplanetary disk can be strongly affected by secular perturbations from a massive body in the system. 
During this phase of planetesimal accretion, the nebula gas is still present in the disk.
Should there be a massive body (e.g., a stellar companion or a planet) perturbing the disk, the coupled effect of secular perturbation (particularly when the perturber is eccentric) and nebula gas drag can lower the relative velocity of planetesimals by aligning the pericenters of their orbits under certain conditions.

A major application of such an effect is on the dynamics of planetesimals in multiple star systems.
The alignment of planetesimal orbits induced by gas drag together with secular perturbation has been confirmed by many studies.  
Using the system configuration of $\alpha$ Centauri, \citet{Marzari_Scholl_2000} found that the gas drag combined with secular perturbation from the secondary star forces a strong alignment of the pericenters of planetesimals orbiting around the primary star. 
In addition to the alignment of orbits, planetesimals of the same size also have similar values of eccentricity because of similar behavior in response to gas drag.
Such a "phasing" of orbits strongly reduces the contribution of eccentricity to the relative velocities of planetesimals, and facilitates planet formation in certain areas in the disk \citep{Marzari_Scholl_2000}.
However, in spite of the fact that the alignment of orbits reduces the encounter velocities of identical-size planetesimals, the size-dependent phasing of orbital elements increases the encounter velocities between planetesimals of different sizes and would likely prevent accretion of planetesimals with a realistic size distribution.
For example, \citet{THEBAULT2006193} further explored the dynamical evolution of planetesimals in more general binary star system configurations by varying the binary separation $a_{\rm{b}}$, eccentricity $e_{\rm{b}}$, and mass ratio.
They found that the erosion threshold is reached in a wide binary parameter space for small planetesimals ($\lesssim 10~$km) but in a more limited region for bigger objects ($\gtrsim 50~$km) \citep{THEBAULT2006193}. 
Studies focusing on specific system configurations (e.g., $\alpha$ Cen A,B, HD 196885) have also shown that planetesimal accretion is challenging for a wide range of parameters \citep[e.g.,][]{Thebault2008,Thebault2009,Thebault2011,Xie_2009}.
Similar points have been made by studies on the dynamical evolution of planetesimals in circumbinary disks \citep[e.g.,][]{Scholl_et_al_2007,Marzari_et_al_2008,Meschiari_2012,Paardekooper2012}.

Another possible type of perturber, apart from stellar companions, is a giant planet (embryo) embedded in the disk.
Considering the effects of gas drag and the gravitational perturbations of a proto-Jupiter, \citet{Marzari1997} studied the orbital evolution of planetesimals in the primordial asteroid belt (2:1 mean motion resonance region).
Collisional evolution of planetesimals in a gas-free disk perturbed by a giant planet embryo has been investigated by e.g. \citet{Thebault_1998}, \citet{Charnoz_et_al_2001}, and \citet{Charnoz2001}.
It has been identified that the collisions of planetesimals act as a diffusion mechanism that propagates and redistributes the strong gravitational perturbations at the resonant regions far away to other disk locations \citep{Thebault_1998}.
Assuming that Jupiter and Saturn has possibly formed via gravitational instability in the cool outer disk and that the terrestrial planets have formed via conventional core accretion method, a notion inspired by disk models in \citet{Boss_1996},  \citet{KW00} and \citet{Kortenkamp_et_al_2001} investigated the consequences of the early formation of these two giant planets. 
Considering gas drag and perturbations from the giant planets, they identified the "Type II" runaway growth mode, where the effects of dynamical friction in the classical "Type I" runaway growth are mimicked by the size dependent phasing of orbital elements \citep{Kortenkamp_et_al_2001}. 

Inspired by the previous studies, we became interested in the consequence of an early formed giant planet on subsequent planetesimal accretion in the external disk. 
Assuming that Jupiter has formed early, we investigated the effect of its secular perturbation together with gas drag on the planetesimals orbiting exterior to Jupiter \citep[][hereafter GK21]{Guo_Kokubo_2021}. 
We found that because of resonance trapping, there are several locations in the external disk at which planetesimals pile up and have chaotic orbital directions with high eccentricities.
However, between these chaotic locations near the mean motion resonances (MMR), we could identify a few dynamically quiet areas where the planetesimal orbits are well aligned. 
At these locations, the relative velocities of planetesimals (particularly for similar-sized bodies) can be reduced, resulting in faster accretion (GK21).

Despite of the abundance of literature related to this topic, the dependence of such an effect of orbital alignment and the consequential reduction of the relative velocity on the parameters of the planet has not been extensively investigated. 
As mentioned earlier, some studies have shown the dependence of the planetesimal relative velocity on binary parameters, including binary separation, mass ratio, and eccentricity \citep[e.g.,][]{THEBAULT2006193, Scholl_et_al_2007}.
However, due to the different system configurations of a binary star system and a single star system with an embedded planet (i.e., the reference frame, the perturbing functions, the range of mass ratio, etc.), the results for binary systems cannot be directly applied to single star systems perturbed by a planet.

To quantitatively determine the effect of the planet parameters, we conduct numerical experiments for different models, each with a planet of a given set of parameters, mass $M_{\rm{p}}$ and eccentricity $e_{\rm{p}}$.
Following GK21, we explore the effect of a giant planet embedded in the disk on the orbital evolution of planetesimals in an external disk with nebula gas in presence.
Although the perturbation from the planet is sensitive to its semi-major axis, we currently fix the planet at 5.2 au (location of Jupiter) since we consider a wide planetesimal disk such that the semi-major axis dependence can be inferred from the distribution of planetesimal orbital elements along the semi-major axis.
In particular, we focus on the relative velocity of planetesimals and its dependence on the mass and eccentricity of the planet.

The remainder of this paper is arranged as follows: in Section \ref{sec:method} we introduce the model setup and the calculation of the relative velocity of planetesimals; in Section \ref{sec:results} we present and explain the dependence of the planetesimal relative velocity on the mass and eccentricity of the planet as well as on planetesimal sizes; we discuss on the implications of our results and the limitations of our models in Section \ref{sec:discussion}; finally, we summarize the paper with a conclusion section.

\section{Method} \label{sec:method}

We numerically calculate the orbital evolution of planetesimals under the coupled effect of nebula gas drag and a giant planet.
We investigate the dependence of the relative velocities of planetesimals on the mass and eccentricity of the planet by varying these two parameters in our models.
The numerical integrator we use is a fourth-order Hermite integrator \citep{Kokubo_Makino_2004}.
More details of the integrator are provided in Section 2.2 in GK21.






\subsection{Model Setup} \label{subsec:model}


The system configuration is the same as that in GK21, i.e., a system with a central star of mass $M_* = 1\, M_{\odot}$, a giant planet embedded at 5.2 au, and a protoplanetary disk comprised of gas and planetesimals.
The gas component follows a power law distribution, with the spatial density at the disk midplane being
\begin{equation}
    \rho_{\rm{g}} = 2.0 \times 10^{-9} \left( \frac{a}{1\,{\rm{au}}}\right)^{-11/4} \rm{g}\, \rm{cm}^{-3}.
    \label{eq:gas_density}
\end{equation}
which corresponds to a 50\% more massive disk than the minimum mass solar nebula (MMSN) \citep{Hayashi_1981}.
Here $a$ is the distance from the central star.

It is worth noting that the assumption of an axisymmetric circular gas disk in our model might not be very accurate because the gas disk is also perturbed by the gravity of the embedded giant planet (see Section \ref{subsec:limitations} for further discussion on the simplification of the gas profile). 
However, there is still no consensus on how to constrain the gas disk reacting against the perturbation from an embedded planet with realistic parameters. 
Therefore, we first approach the problem with a simplest case, i.e. an axisymmetric gas disk, in the purpose of finding out the basic effect of the perturbation from a planet along with nebula gas drag on the orbital evolution of planetesimals.

The planetesimals are distributed in an annulus of 8-15 au, following a power law surface density distribution $\Sigma_{\rm{d}} \propto a^{-3/2}$. 
The mass spectrum of the planetesimals is [$10^{16}, 10^{17}, 10^{18}, 10^{19}, 10^{20}$] g.
For the selection of planetesimal mass range, we follow GK21 and cut the smallest masses $m=10^{13},\quad 10^{14},$ and $10^{15}~$g. 
The reason for cutting these smaller particle masses is that these particles drift inward fast as a result of strong coupling to gas drag, making them less efficient targets for the purpose of demonstrating the relative velocities between these small particles and other larger planetesimals.
Meanwhile, since we focus on the dependence on the planet mass and eccentricity in this paper, we simplify the size distribution of planetesimals and emphasize on the variation of planetesimal velocities as we adjust the planet parameters.

For each mass, we have $N = 10,000$ particles.
The initial orbital eccentricity and inclination of planetesimals follow the Rayleigh distribution with dispersion $\sigma_e = \sigma_i = 2 r_{\rm{H}}/a$, where the Hill radius of the planetesimal $r_{\rm{H}}$ is given by Eq. (4) in GK21.
The other orbital elements $(\Omega, \omega, \tau)$ are randomly selected from 0 to $2\pi$.
The density of the planetesimals is $1\, \rm{g} \, \rm{cm}^{-3}$ since they are distributed outside the snow line. 

The gas drag force is calculated from the formula given in \citet{Adachi_et_al_1976}:
\begin{equation}
    f_{\rm{g}} = \frac{1}{2}C_{\rm{D}} \pi r^2 \rho_{\rm{g}} u^2.
    \label{eq:gas_drag}
\end{equation}
The value of $C_{\rm{D}}$ and the drag law are described in Section 2.1 in GK21.
To estimate the timescale of gas damping on planetesimal eccentricity, we use the equation given in \citet{kokubo2012dynamics}
\begin{eqnarray}
    \tau_{\rm{gas}} = \frac{e}{|{\rm{d}}e/{\rm{d}}t|} \sim 10^5 \left(\frac{e_{\rm{f}}}{0.1}\right)^{-1} \left(\frac{m}{M_{\oplus}}\right)^{1/3} \nonumber\\
    \left(\frac{\rho}{3\, {\rm{g}}\,{\rm{cm}}^{-3}}\right)^{2/3} \left(\frac{a}{1\,{\rm{au}}} \right)^{13/4}\, \rm{yr}.
    \label{eq:tau_damp_out}
\end{eqnarray}
For the outer orbits, we use the value of the forced eccentricity $e_{\rm{f}}$ because in the outer disk, where the surface density of gas is low and thus the gas drag is weak, the eccentricities of planetesimals are close to the forced eccentricity, particularly for larger-mass particles.

Following GK21, we follow the system evolution for $t\simeq 3~$Myr, which roughly corresponds to the gas damping timescale $\tau_{\rm{gas}}$ for the largest planetesimal ($m = 10^{20}~$g) near 14 au (see Fig. 3 in GK21).

The equation of motion of a planetesimal is given by 
\begin{eqnarray}
\odv{\boldsymbol{v}_i}{t} = &-& GM_* \frac{\boldsymbol{x}_i}{|\boldsymbol{x}_i|^3} + G M_{\rm{p}} \frac{\boldsymbol{x}_{\rm{p}}-\boldsymbol{x}_i}{|\boldsymbol{x}_{\rm{p}} - \boldsymbol{x}_i|^3} \nonumber \\ 
&-& GM_{\rm{p}} \frac{\boldsymbol{x}_{\rm{p}}}{|\boldsymbol{x}_{\rm{p}}|^3} + \boldsymbol{f}_{\rm{g}}.
\label{eq:eom}
\end{eqnarray}
The subscript "p" denotes parameters of the planet.
The third term is the indirect term, which arises from the indirect perturbation of the planet on the central star.
Following GK21, we currently neglect the mutual gravitational interaction between planetesimals for simplicity. 

\subsection{Planet parameters} \label{subsec:planet_parameters}

In order to determine the dependence of the dynamical behavior (particularly the relative velocities) of planetesimals on the planet parameters, we vary the mass and eccentricity of the planet in each model.
We consider four values each for the planet mass and eccentricity, roughly evenly spaced on a log scale: $M_{\rm{p}} = 0.1$, 0.3, 1, 3$\, M_{\rm{J}}$ and $e_{\rm{p}} = 0.01$, 0.02, 0.05, 0.1.
The case in which the planet parameters equal to those of Jupiter, namely when $M_{\rm{p}} = 1\, M_{\rm{Jup}}$ and $e_{\rm{p}} = 0.05$ is set as the fiducial model.
When we vary the planet mass $M_{\rm{p}}$, we fix the planet eccentricity $e_{\rm{p}} = 0.05$; and when we vary $e_{\rm{p}}$, we fix $M_{\rm{p}} = 1\, M_{\rm{Jup}}$. 
All the other settings remain the same.
Therefore, we have seven models in total.
Table \ref{table:model_names} lists the model names and planet parameters for each model.

\begin{deluxetable*}{Lccccccc}
\label{table:model_names}
\tablecaption{Planet models}
\tablehead{
\colhead{Model} & \colhead{M1E3} & \colhead{M2E3} & \colhead{M4E3} & \colhead{M3E3} & \colhead{M3E1} & \colhead{M3E2} & \colhead{M3E4}
}
\startdata
M_{\rm{p}}~(M_{\rm{Jup}}) & 0.1 & 0.3 & 3 & 1 & 1 & 1 & 1 \\
e_{\rm{p}} & 0.05 & 0.05 & 0.05 & 0.05 & 0.01 & 0.02 & 0.1 \\
\enddata
\tablecomments{The name of each model and the planet mass and eccentricity in each model.}
\end{deluxetable*}

Although we assume that the planet perturbing the disk has formed early, we do not require that it has formed via gravitational instability (G.I.), as in \citet{KW00} and \citet{Kortenkamp_et_al_2001}. 
These two papers considered a scenario in which both Jupiter and Saturn have formed prior to the accretion of solid material in the inner Solar System.
Such a system configuration requires the Jupiter and Saturn have formed very early, possibly via the \textit{disk instability} mechanism (or G.I.), which predicts that giant planets form through the clumping and collapse of the gaseous protoplanetary disk \citep{Boss_1997}.
Although the \textit{disk instability} model provides a good route to form giant planets on very short timescales, it is less favored for explaining the formation of the gas giants in the Solar System.
Instead, it is a more preferable mechanism for producing massive gas giants on wide orbits \citep[e.g.,][]{BOLEY2010509,Boss_2011,Fletcher_2019}.

Our paper does not require that the existing planet has formed through G.I. and is compatible with the conventional \textit{core accretion} model.
The \textit{core accretion} model predicts that giant planets form from the inside out \citep[e.g.,][]{Hayashi_1985, Pollack_et_al_1996}.
Therefore, our assumption that the planet in our models have formed earlier than possible planetary cores exterior to its orbit is natural.

In addition, by varying the mass and eccentricity of the perturbing planet, we loosen the constraint on its route of formation.
Specifically, we consider two planet masses smaller than that of the present-day Jupiter to account for a possibly smaller young gas giant which has formed through core accretion, and subsequently perturbs the solid materials in the disk. 
By varying the planet eccentricity, we account for the uncertainty of the eccentricity of the young gas giant, as the eccentricities of planets in multi-planet systems are unlikely to maintain constant through time.
It has been shown that the eccentricity of Jupiter oscillates periodically as a result of secular interaction with other planets in the Solar System \citep{Brouwer_vanWoerkom_1950}.

Because we place our interests on an early phase of planet formation when only one young gas giant is present in the disk, we do not consider the possible migration of the planet.
Some studies have pointed out that the giant planets in the Solar System might have significantly migrated at different stages in their history \citep[e.g.,][]{Walsh2011,Liu2022}.
However, the migration of planets shown in these studies takes place at a later stage when multiple giant planets have formed (or partially formed): the Grand Tack model demands that Saturn migrates inward faster than Jupiter and becomes trapped in the 2:3 resonance so that the direction of migration of both planets is then reversed \citet{Walsh2011}; the instability of giant planets triggered by disk dispersal happens at an even later stage when the nebula gas starts to dissipate \citep{Liu2022}.
Therefore, the migration of the planet is out of the scope of the current study, which focuses on the early dynamical evolution of planetesimals before the formation of a second giant planet core.
The possible migration of the existing giant planet, however, might be important during the accretion of a protoplanet once it is formed, as the dynamical behavior of the planetesimals sensitively depends on the distance from the perturber.

\subsection{Relative velocity} \label{subsec:method_velocity}

Because the relative velocity $\langle \Delta v \rangle$ oscillates over time, and we are interested in its dependence on planet parameters instead of its time evolution, we calculate the time-averaged value of $\langle \Delta v \rangle$ for a given pair of planetesimal masses instead of the instantaneous value in a certain snapshot.
This time-averaged value of $\langle \Delta v \rangle$ should be representative of the model setup and independent of the initial conditions.
As we have discussed in Section 3 in GK21, under the coupled effect of secular perturbation and gas drag, the eccentricity vectors of a planetesimal population circulate the forced eccentricity with shrinking radius (i.e. proper eccentricity) on the eccentricity vector ($e_x$-$e_y$) plane until equilibrium is reached.
Therefore, along the axis of time $t$, the value of $\langle \Delta v \rangle$ (particularly for large planetesimal mass ratios) oscillates around a certain value with a generally decreasing oscillation amplitude and absolute magnitude.
Since the initial eccentricities and their dispersion are very small, we avoid the initial values and calculate the time-averaged velocities starting from a time when the eccentricity vectors have circled once around the forced eccentricity, i.e., after the secular timescale $\tau_{\rm{sec}}$ (see Eq. (26) in GK21).
Because $\tau_{\rm{sec}}$ depends on the planet parameters (mainly on planet mass), we list the times of snapshots which are chosen as the closest values to $\tau_{\rm{sec}}$ for each model in Table \ref{table:time_span}.
The time-averaged velocities of planetesimals are calculated using the snapshots starting from these times to the end of the simulations ($t = 2985070~$yr).


\begin{deluxetable}{ccc}
\label{table:time_span}
\tablecaption{Starting time for calculating the average of the relative velocity (unit: year)}
\tablehead{
\colhead{Model} & \colhead{10 au} & \colhead{12 au}
}
\startdata
M1E3 & 1545036 & 2925068\\
M2E3 & 510012 & 975022\\
M3E3 & 150003 & 285006\\
M4E3 & 45001 & 90002\\
M3E1 & 150003 & 285006\\
M3E2 & 150003 & 285006\\
M3E4 & 150003 & 285006\\
\enddata
\tablecomments{The starting times of snapshots to calculate the time-averaged relative velocities for each model. The values are chosen to be the nearest to the corresponding $\tau_{\rm{sec}}$ of each model. When the value of $\tau_{\rm{sec}}$ falls in the interval of two snapshots, we choose the earlier snapshot to start the calculation.}
\end{deluxetable}

To provide a more conservative estimate of the relative velocities, we made some slight improvements to the calculation method of $\langle \Delta v \rangle$ compared with that in GK21.
For a population of identical-mass planetesimals in a semi-major axis bin of width $\Delta a$, the relative velocity is the velocity dispersion of this population ($\langle \Delta v \rangle = \sigma_v$), and the calculation method remains the same as in our previous paper (i.e. $\sigma_v \simeq \sigma_e v_{\rm{K}}$).
$v_{\rm{K}} = a \sqrt{\frac{G M_*}{a^3}}$ is the Keplerian velocity.
$\sigma_e$ is the dispersion of the eccentricity vectors of a population:
\begin{eqnarray}
    \sigma_e & = & \sqrt{\sigma_{e_x}^2 + \sigma_{e_y}^2} \nonumber\\
             & = & \sqrt{\frac{\Sigma^N_{i=1} (e_{x,i} - \bar{e}_x)^2}{N} + \frac{\Sigma^N_{i=1} (e_{y,i} - \bar{e}_y)^2}{N}}
\end{eqnarray}
For different-mass particles that fall into the same semi-major axis bin, we fix the target mass and vary the projectile mass and thus the mass ratio.
The target mass is always larger than the projectile mass.
The relative velocity between populations with different masses is given by 
\begin{equation}
    \langle \Delta v \rangle \simeq v_{\rm{K}} \cdot (\Delta e + \sigma_{e,{\rm{tar}}}),
    \label{eq:relative_velocity}
\end{equation}
where $\Delta e$ is the difference between the means of the eccentricity vector distributions of the two populations:
\begin{equation}
    \Delta e = \sqrt{(\bar{e}_{x,\rm{tar}} - \bar{e}_{x,\rm{pro}})^2 + (\bar{e}_{y,\rm{tar}} - \bar{e}_{y,\rm{pro}})^2}.
    \label{eq:delta_e}
\end{equation}

Using Eq. (\ref{eq:relative_velocity}) we can account for the large dispersion of the eccentricity vectors of the target planetesimals due to weak response to gas drag while calculating the relative velocity between the two populations.
The resulting velocities depend on the semi-major axis bin width $\Delta a$, because $\Delta a$ reflects the extent of the differential precession of the eccentricity vectors on the $e_x$-$e_y$ plane.
Larger $\Delta a$ leads to a larger number of particles for calculating the eccentricity dispersion as well as more sufficient differential precession of the eccentricity vectors of these particles (GK21).
Such effects would result in larger values of planetesimal velocities. 
Following GK21, we set $\Delta a = 0.1~$au for calculating the relative velocities (see Section 2.4 in GK21).
Such a bin width provides enough particles for us to calculate the dispersion and reflects well the semi-major axis dependence of the velocities.
For the purpose of qualitatively showing the dependence of the planetesimal velocities on the planet parameters, we do not stress the importance of the choice of $\Delta a$ here.
However, for a quantitative comparison with the erosion threshold velocities, our statistical method of velocity calculation might be subjected to inaccuracy.
We discuss the limitation of our method of velocity calculation in Section \ref{subsec:velocity_uncertainty}.

\section{Results} \label{sec:results}

Owing to a smaller gas surface density compared with that of the inner disk, most of the outer orbits of planetesimals in our models have not reached equilibrium within the integration time.
This is particularly observed for particles of larger masses, e.g., $m\gtrsim 10^{18}\,$g.
Consequently, the time evolution of the distribution of eccentricity vectors of the outer orbits can generally be described as circulating around the forced eccentricity on the $e_x$-$e_y$ plane.
Here, we focus on the impact of varying the mass and eccentricity of the planet.
For more details on the dynamical evolution of the outer orbits under the coupled effect of secular perturbation and gas drag, we refer to GK21.

\subsection{Dependence on planet mass}
\begin{figure*}
    \centering
    \includegraphics[width=0.8\textwidth]{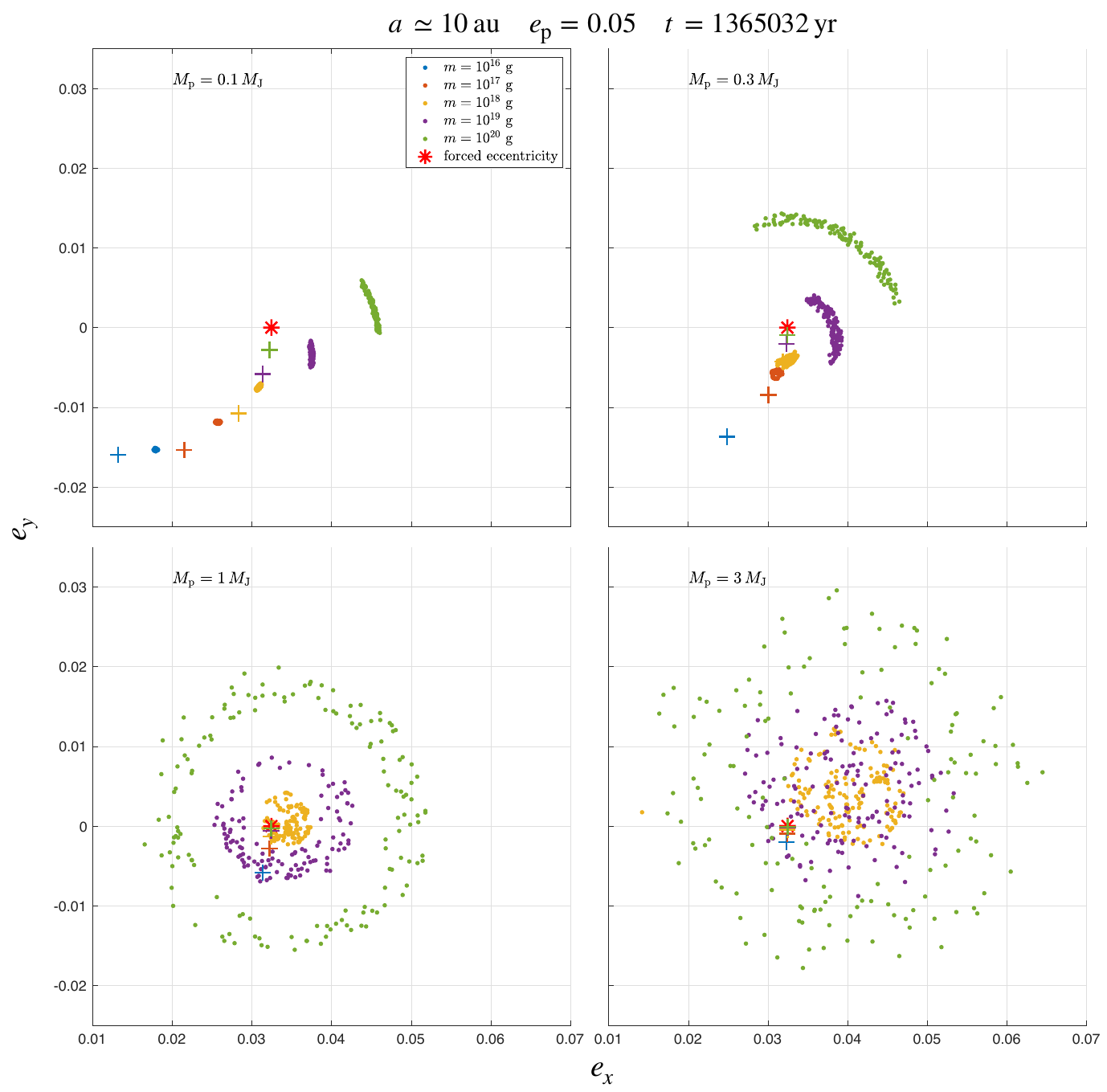}
    \caption{Distribution of the eccentricity vectors of particles near 10 au for four planet masses (snapshots taken at $t = 1365032~$yr). The colors represent different planetesimal masses. The plus signs $+$ in each panel represent the equilibrium locus of eccentricity vector for different-size particles.}
    \label{fig:eccvec_10au_Mp}
\end{figure*}

The increase in planet mass affects the alignment of the orbits from two aspects:

\noindent \textit{Short-term effect}: as $M_{\rm{p}}$ increases, the gravitational kick during each encounter with the planet becomes stronger. 
As a result, the dispersion of each population of identical mass increases on the $e_x$-$e_y$ plane. 
Fig. \ref{fig:eccvec_10au_Mp} shows the distribution of eccentricity vectors near 10 au at $t = 1365032\,$yr. 
The time of snapshots is selected such that the results reflect the orbits at the typical time of core formation ($\sim 1\,$Myr) and the distributions of each population are clearly identified on the $e_x$-$e_y$ plane.
The increase in eccentricity dispersion $\sigma_e$ can be observed from the increasing radius of each population (if in a circular shape) or the larger width of the arc-like distributions (i.e. when differential precession is prominent). 
This increase in $\sigma_e$ directly influences the velocity dispersion $\sigma_v$ of identical-mass particles (see Section \ref{subsec:method_velocity}).
    
\noindent \textit{Long-term (secular) effect}: for larger planet mass, the secular perturbation becomes stronger.
A straightforward effect is that the secular perturbation timescale $\tau_{\rm{sec}}$ shrinks as $M_{\rm{p}}$ increases (GK21). 
With a shorter $\tau_{\rm{sec}}$, the differential precession of eccentricity vectors is more sufficient.
This can be visualized from the longer arc-like distributions in the upper right panel compared with those in the upper left panel in Fig. \ref{fig:eccvec_10au_Mp}. 
For even larger planet masses ($M_{\rm{p}} = 1\, M_{\rm{J}}$ and $M_{\rm{p}} = 3\, M_{\rm{p}}$), the distributions of $\boldsymbol{e}$ of the large-mass particles become a complete circle around the forced eccentricity, indicating strong secular perturbation and weak gas damping.
Such an effect also results in a larger $\sigma_e$ and thus a larger $\sigma_v$ of identical-mass particles. 

It is worth noting that the more compact distribution of the equilibrium locus of $\boldsymbol{e}$ as $M_{\rm{p}}$ increases (as shown by the plus signs in Fig. \ref{fig:eccvec_10au_Mp}) does not lead to lower relative velocity between different populations as in the inner orbits. 
This is primarily because, although the difference between the mean location of two populations $\Delta e$ decreases as the equilibrium locus becomes more compact, in this case the actual relative velocity between two populations $\Delta v$ is dominated by the dispersion $\sigma_e$ of each population, which increases dramatically as $M_{\rm{p}}$ increases.
In other words, as $M_{\rm{p}}$ increases, the increase in $\sigma_e$ overcomes the decrease in $\Delta e$, and dominates the dependence of the relative velocity on the planet mass in the outer orbits. 

\begin{figure*}
    \centering
    \gridline{\fig{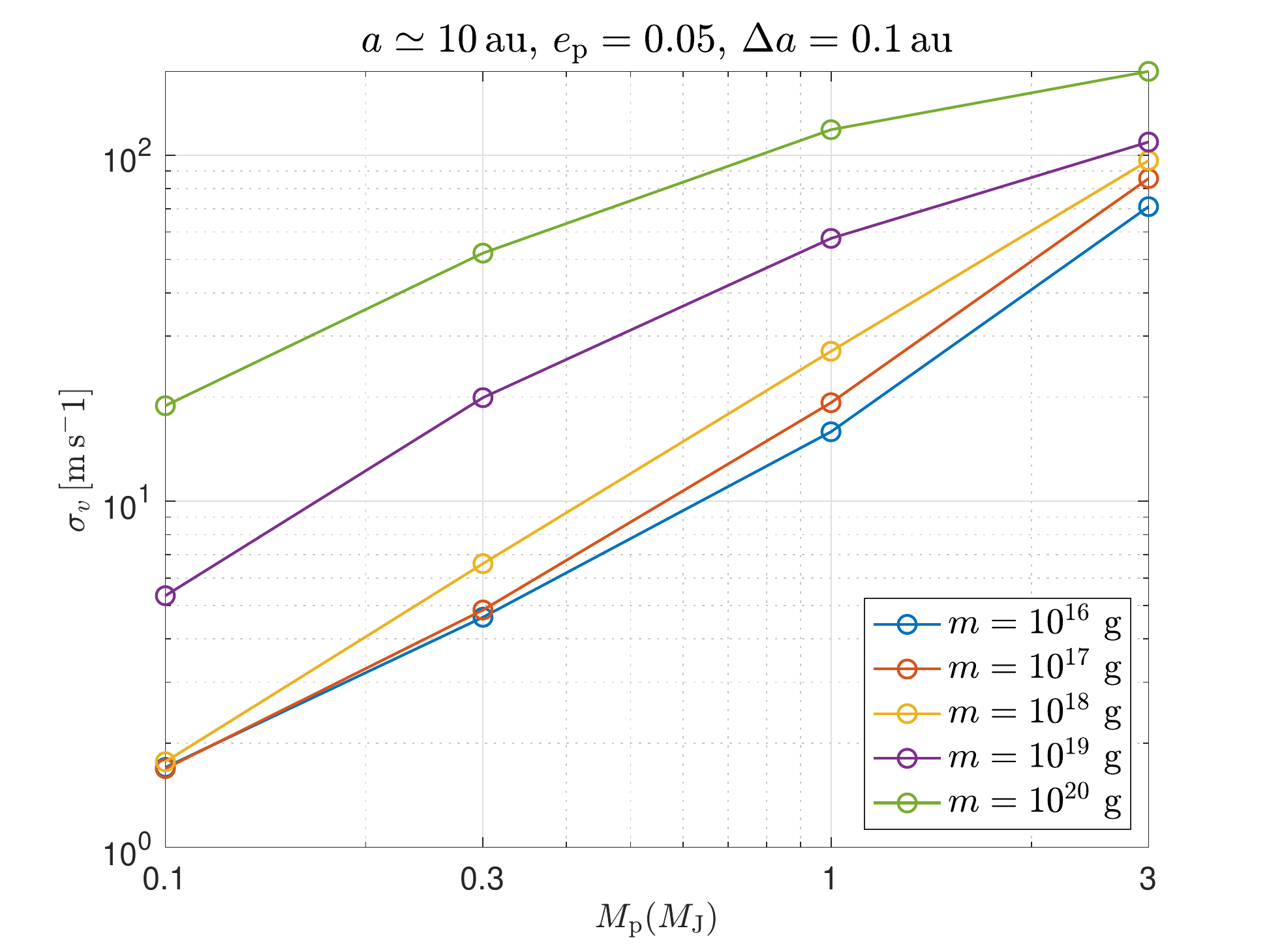}{0.45 \textwidth}{(a)}
             {\fig{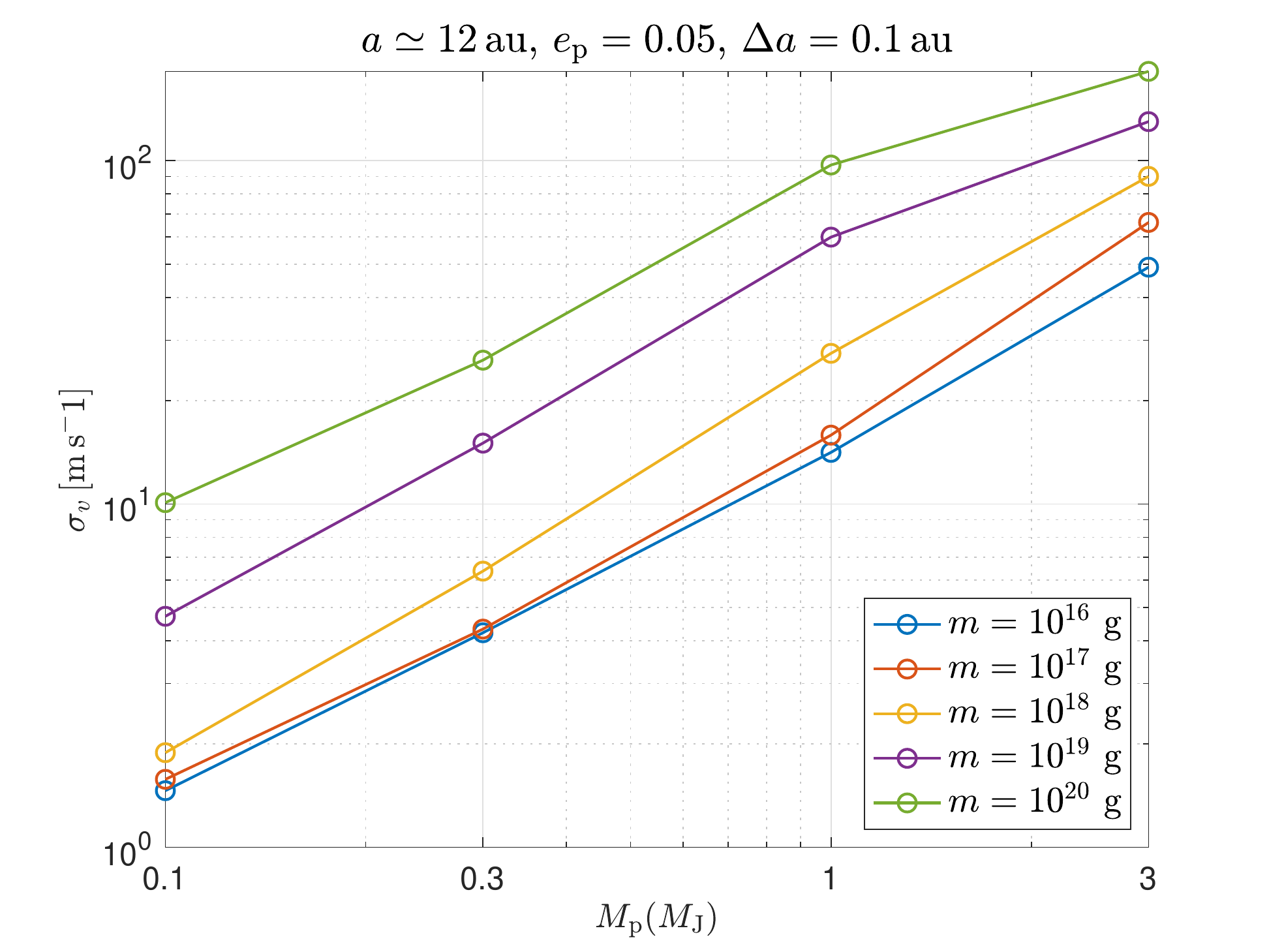}{0.45 \textwidth}{(b)}}}
    \caption{Dependence of the velocity dispersion $\sigma_v$ of identical-mass planetesimals on planet mass near (a) 10 au and (b) 12 au. The semi-major axis bin width $\Delta a$ is 0.1 au.}
    \label{fig:Mp_sigmaV_out}
\end{figure*}

\begin{figure*}
    \centering
    \gridline{\fig{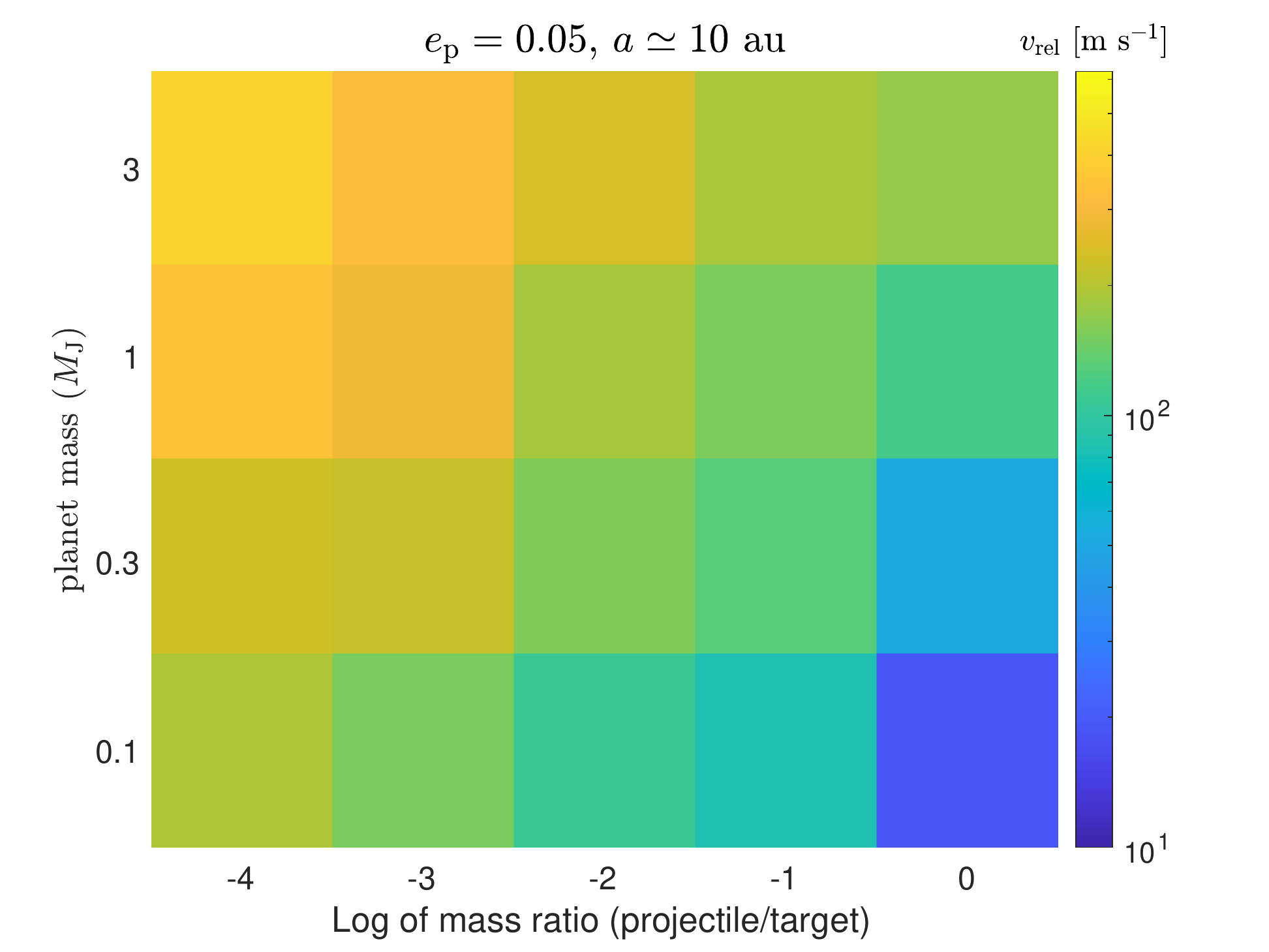}{0.45\textwidth}{(a)}
             {\fig{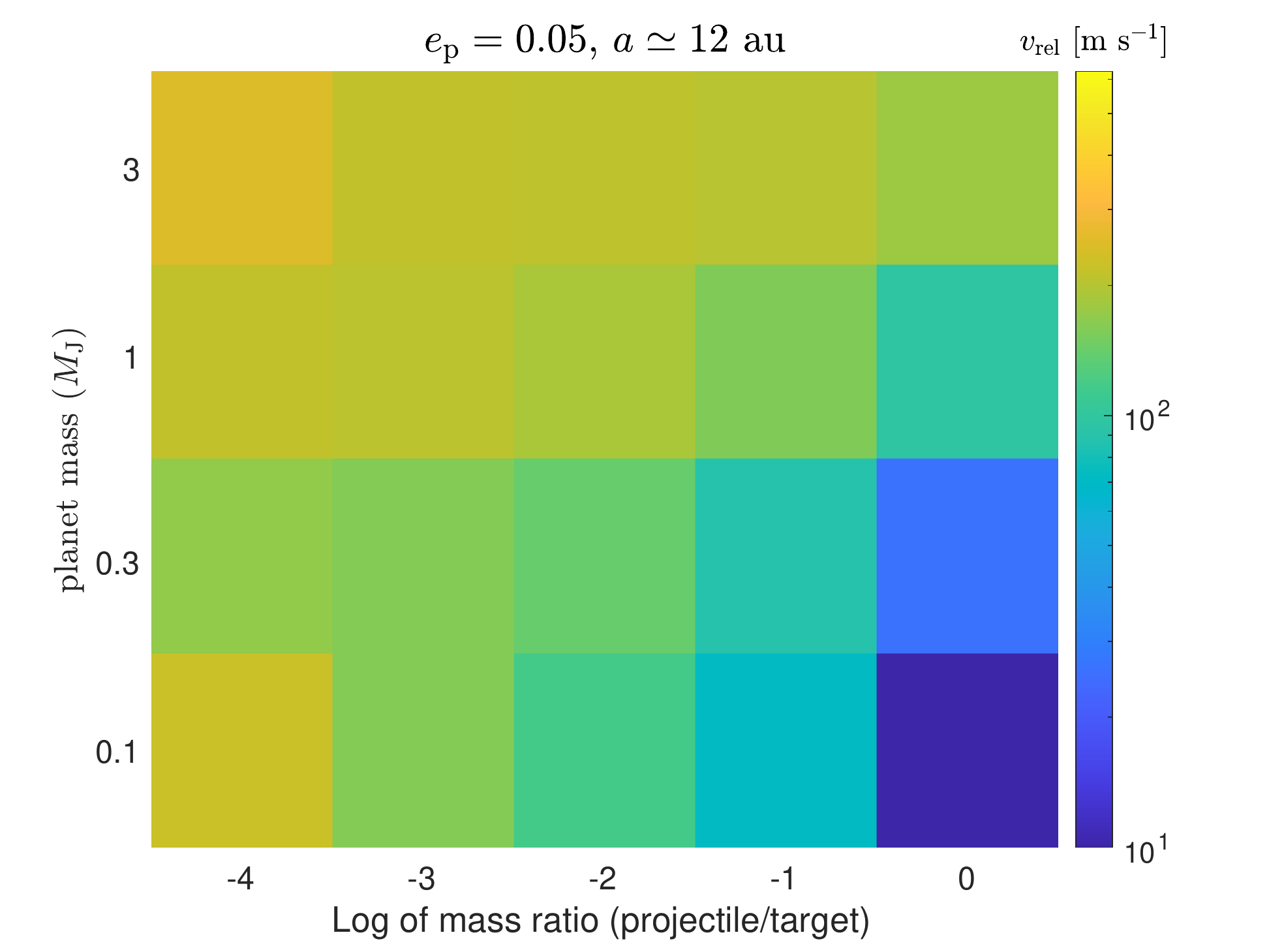}{0.45\textwidth}{(b)}}}
    \caption{Relative velocity between particles of different mass ratio for different planet masses near (a) 10 and (b) 12 au. The color scale indicates the value of the relative velocity of planetesimals. The target mass is fixed as $m = 10^{20}~$g.}
    \label{fig:R-Mp_out}
\end{figure*}

Fig. \ref{fig:Mp_sigmaV_out} shows the dependence of $\sigma_v$ on $M_{\rm{p}}$ near 10 and 12 au.
At both $a \simeq 10\,$au and $a \simeq 12\,$au, $\sigma_v$ increases monotonically as $M_{\rm{p}}$ increases. 
Fig. \ref{fig:R-Mp_out} shows the relative velocities between two populations of different masses near 10 and 12 au. 
For a given particle mass ratio,  $\langle \Delta v \rangle$ increases as the planet mass increases.
In addition, for a given planet mass, $\langle \Delta v \rangle$ increases as particle mass ratio increases.
This trend is the same for both $a \simeq 10\,$au and $a \simeq 12\,$au, with $\langle \Delta v \rangle$ values generally higher at 10 au (GK21).

\subsection{Dependence on planet eccentricity}

The effect of increasing planet eccentricity, compared with that of increasing planet mass, is slightly less straightforward. 
Briefly, the impact of increasing $e_{\rm{p}}$ on $\sigma_v$ lies in the balance between the competing effects of gas damping and secular perturbation.
Meanwhile, increasing $e_{\rm{p}}$ increases $\Delta v$ through a more scattered distribution of the equilibrium locus of $\boldsymbol{e}$ for different particle masses.

\begin{figure*}
    \centering
    \includegraphics[width=0.8\textwidth]{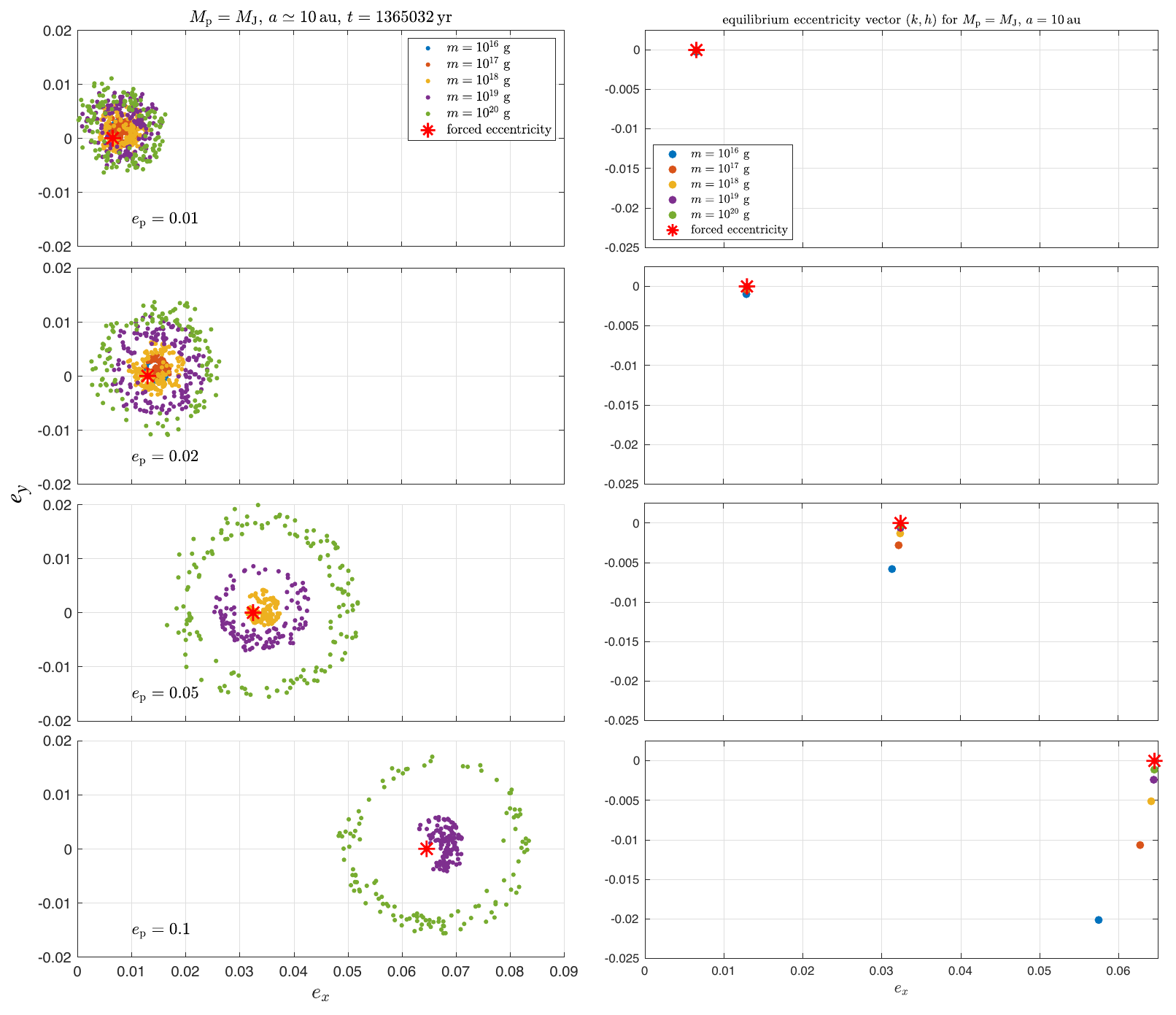}
    \caption{Left column: distribution of eccentricity vectors near 10 au; each panel corresponds to a model with a different planet eccentricity. Right column: equilibrium locus of the eccentricity vector for different-size particles. Each panel corresponds to a model with a different planet eccentricity.}
    \label{fig:eccvec_10au_ep}
\end{figure*}

Fig. \ref{fig:eccvec_10au_ep} shows the distribution of the eccentricity vectors near 10 au at $t = 1365032\,$yr for four different values of $e_{\rm{p}}$. 
The four snapshots in the right column show the corresponding equilibrium loci of $\boldsymbol{e}$ for different particle masses. 
As $e_{\rm{p}}$ increases, the forced eccentricity $e_{\rm{f}}$ increases, and the equilibrium loci of $\boldsymbol{e}$ become more scattered. 
Because the increase in $e_{\rm{p}}$ does not significantly increase the dispersion $\sigma_e$ of each population, the increase in $\Delta e$ as a result of more scattered equilibrium locus leads to a mild increase in the relative velocity $\langle \Delta v \rangle$ between the two populations.
The variation in $\sigma_v$, although not significant, is a combined result of multiple effects, and therefore depends on multiple factors, including planet eccentricity, particle mass, and even the  semi-major axis.

\begin{figure*}
    \centering
    \gridline{\fig{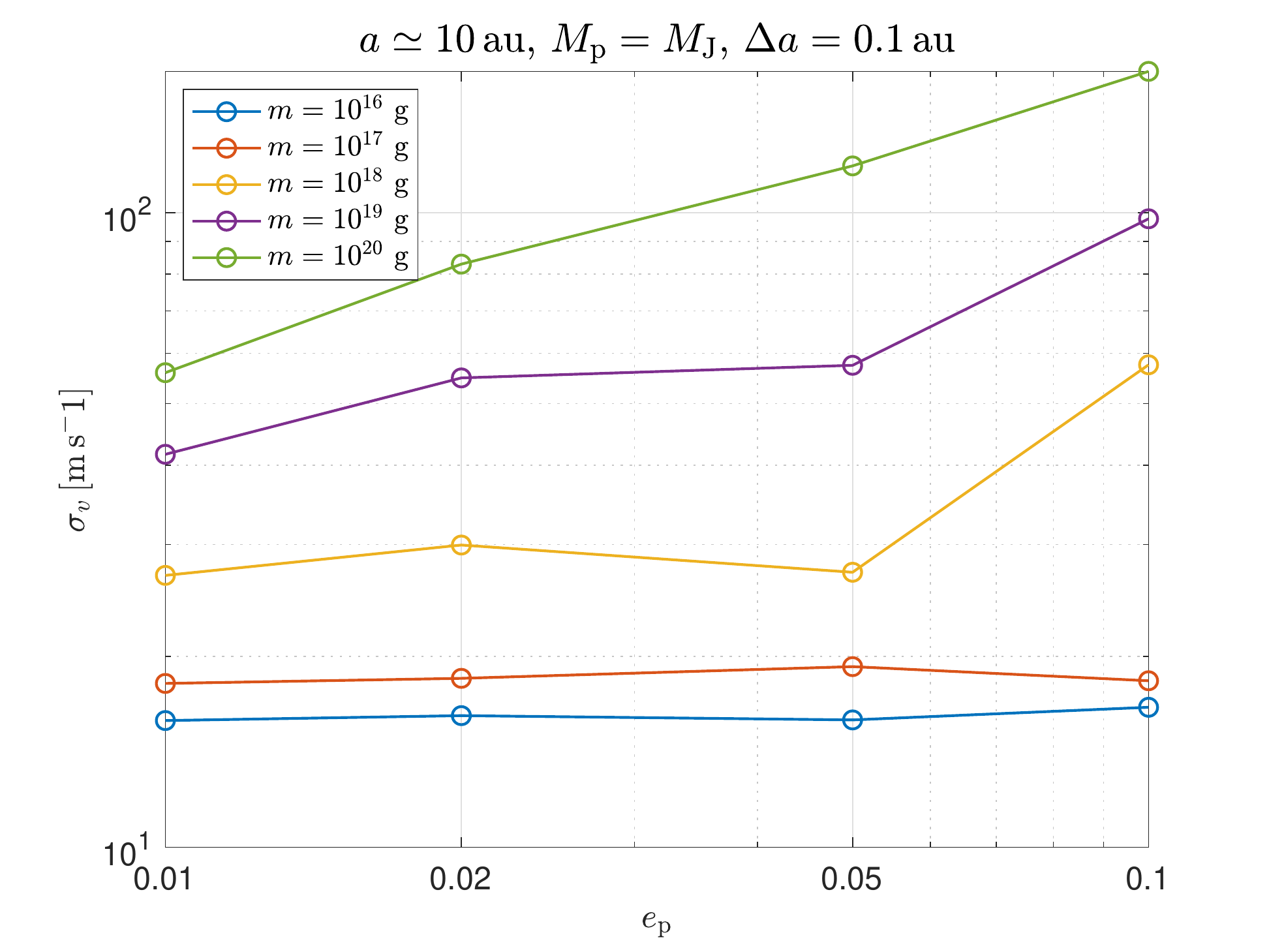}{0.45 \textwidth}{(a)}
             {\fig{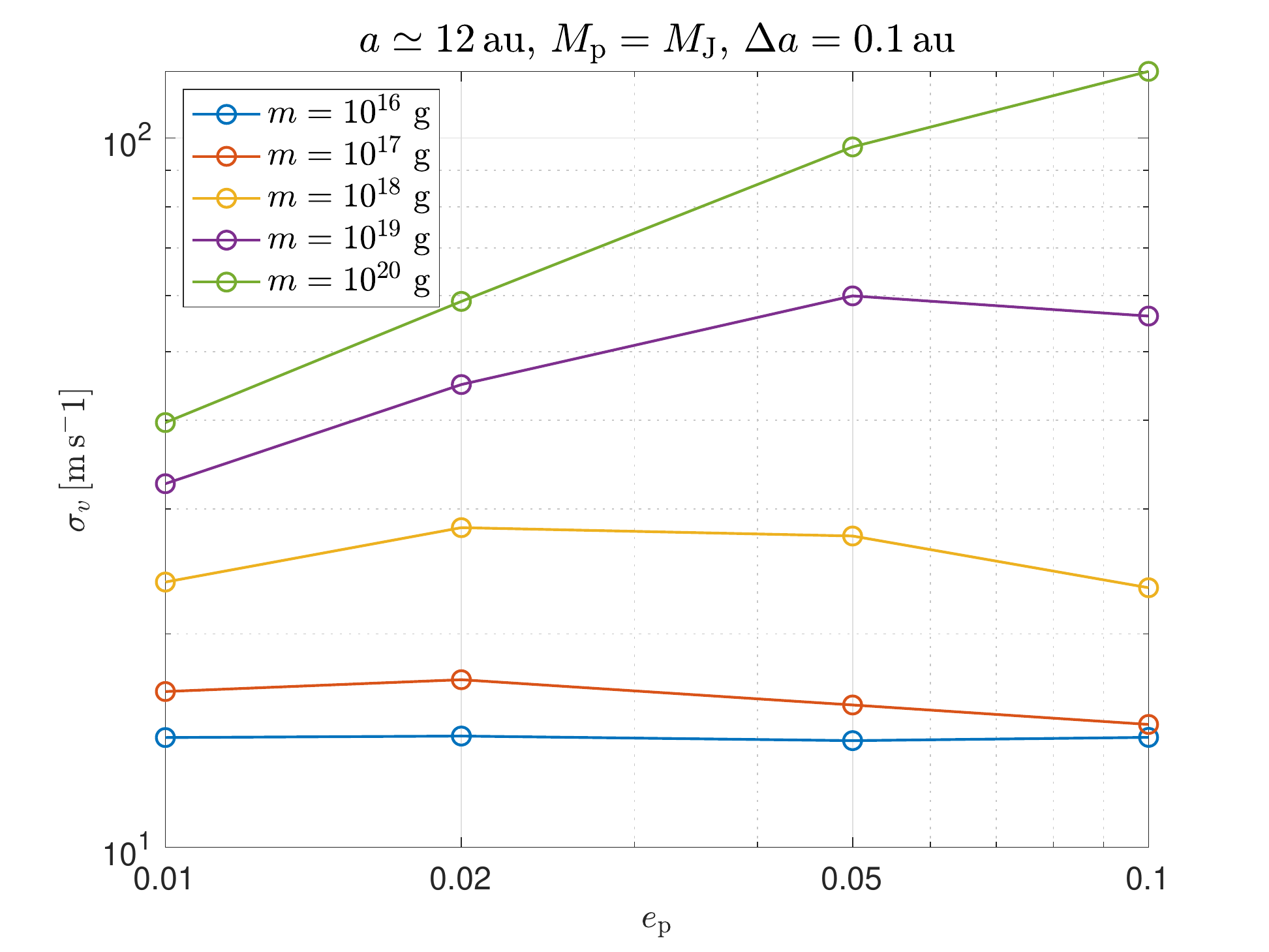}{0.45 \textwidth}{(b)}}}
    \caption{Dependence of velocity dispersion $\sigma_v$ on planet eccentricity near 10 au (a) and 12 au (b).}
    \label{fig:ep_sigmaV_out}
\end{figure*}

\begin{figure*}
    \centering
    \gridline{\fig{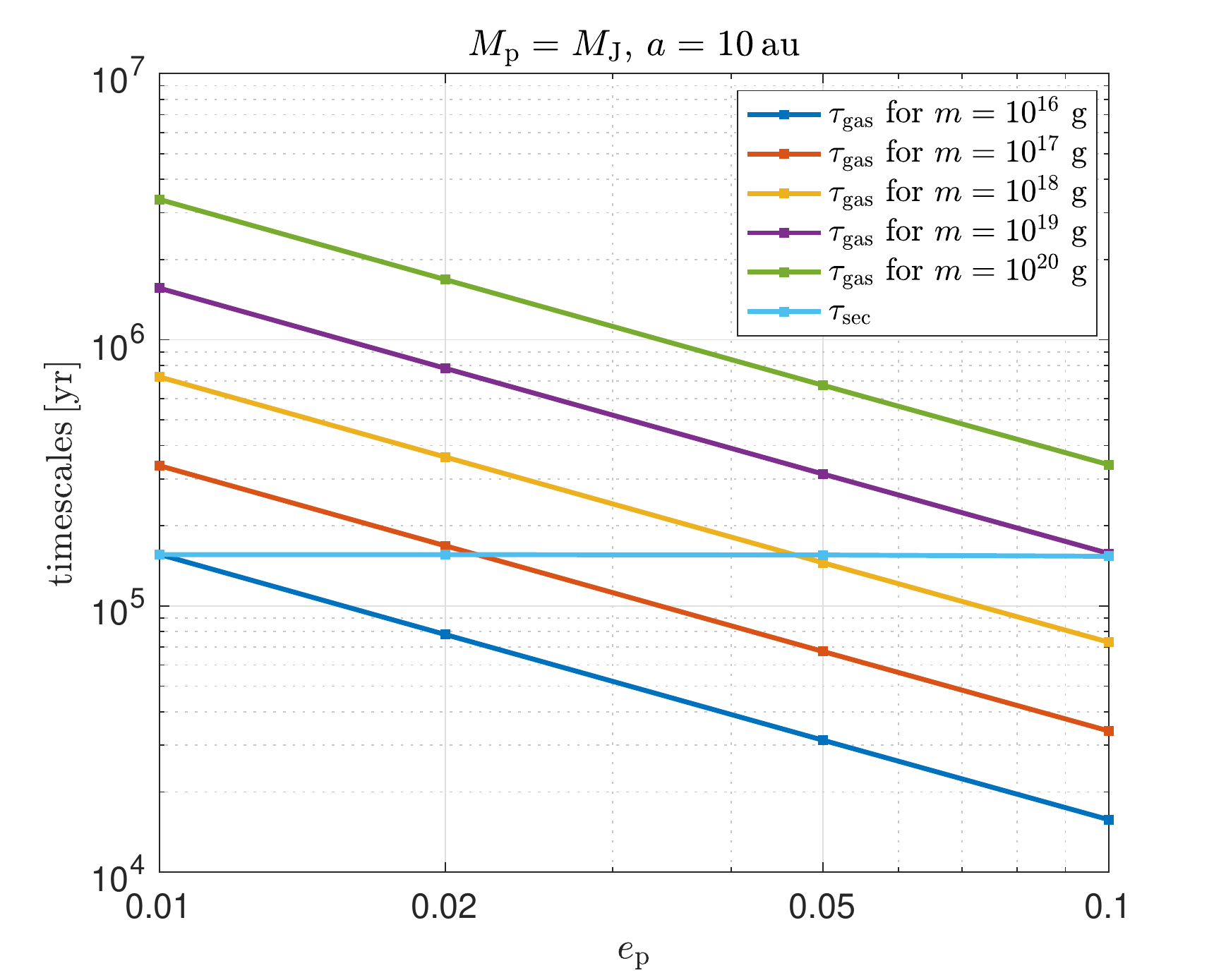}{0.45 \textwidth}{(a)}
             {\fig{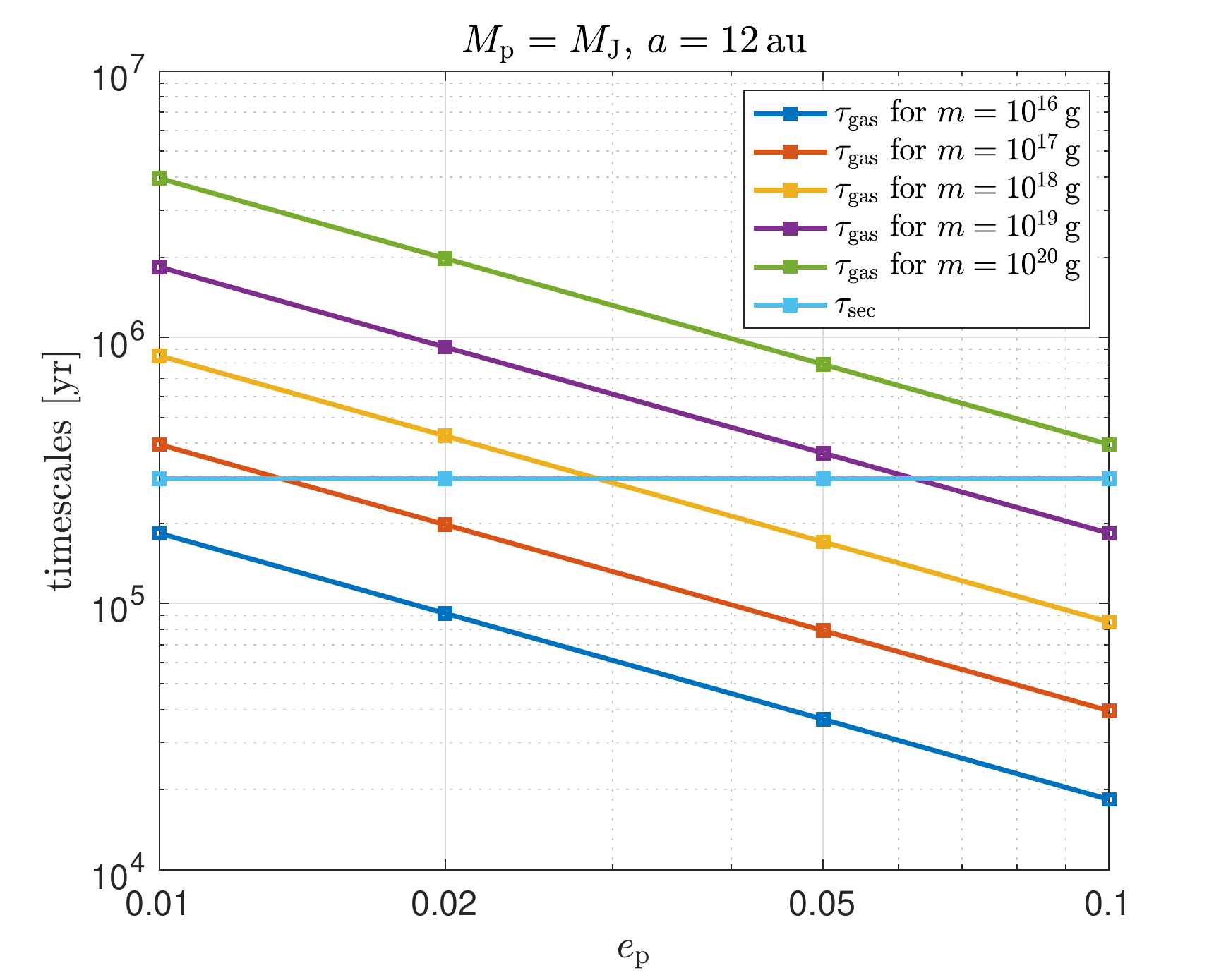}{0.45 \textwidth}{(b)}}}
    \caption{Timescales of gas damping $\tau_{\rm{gas}}$ and secular perturbation $\tau_{\rm{sec}}$ as functions of planet eccentricity near 10 and 12 au. The light blue lines in both plots show the secular timescale $\tau_{\rm{sec}}$, which barely depends on $e_{\rm{p}}$.}
    \label{fig:ep_timescale_out}
\end{figure*}

Fig. \ref{fig:ep_sigmaV_out} shows the dependence of $\sigma_v$ on $e_{\rm{p}}$ near 10 and 12 au.
Unlike the dependence on $M_{\rm{p}}$, the dependence on $e_{\rm{p}}$ is non-monotonic. 
This non-monotonic trend can be explained by the balance between the gas damping and secular perturbation.
Fig. \ref{fig:ep_timescale_out} shows the timescale estimates near 10 and 12 au for each model. 
While the gas damping timescale $\tau_{\rm{gas}}$ decreases with increasing $e_{\rm{p}}$, the secular perturbation timescale $\tau_{\rm{sec}}$ barely varies. 
Separated by the almost flat line of $\tau_{\rm{sec}}$, the upper-left region in both plots in Fig. \ref{fig:ep_timescale_out} is the secular-perturbation-dominated regime, and the lower right region is the gas-damping-dominated regime. 
In other words, when $\tau_{\rm{sec}} < \tau_{\rm{gas}}$, the dynamical behavior of planetesimals is dominated by secular perturbation such that $\sigma_v$ increases with increasing $e_{\rm{p}}$.
In contrast, when $\tau_{\rm{sec}} > \tau_{\rm{gas}}$, the dynamical behavior of planetesimals is dominated by gas damping, so that $\sigma_v$ decreases with increasing $e_{\rm{p}}$.
Meanwhile, because 10 au lies between the 2:1 and 3:1 mean motion resonance, for high planet eccentricity ($e_{\rm{p}} \gtrsim 0.05$), the eccentricity of some large-mass particles ($m \gtrsim 10^{18}~$g) could be excited to high values as a result of the increased resonance width. 
Thus, at $a \simeq 10\,$au, $\sigma_v$ for particle mass $m=10^{18}\,$g decreases at $e_{\rm{p}} = 0.05$ and then dramatically increases at $e_{\rm{p}} = 0.1$. The increasing trend of $\sigma_v$ for particle mass $m=10^{19}\,$g first mitigates when $e_{\rm{p}}$ increases from 0.02 to 0.05, and then becomes steep again when $e_{\rm{p}}$ raises to 0.1.
However, at $a \simeq 12\,$au, a location far from resonance, no such twisting trend of $\sigma_v$ is observed as near 10 au.

\begin{figure*}
    \centering
    \gridline{\fig{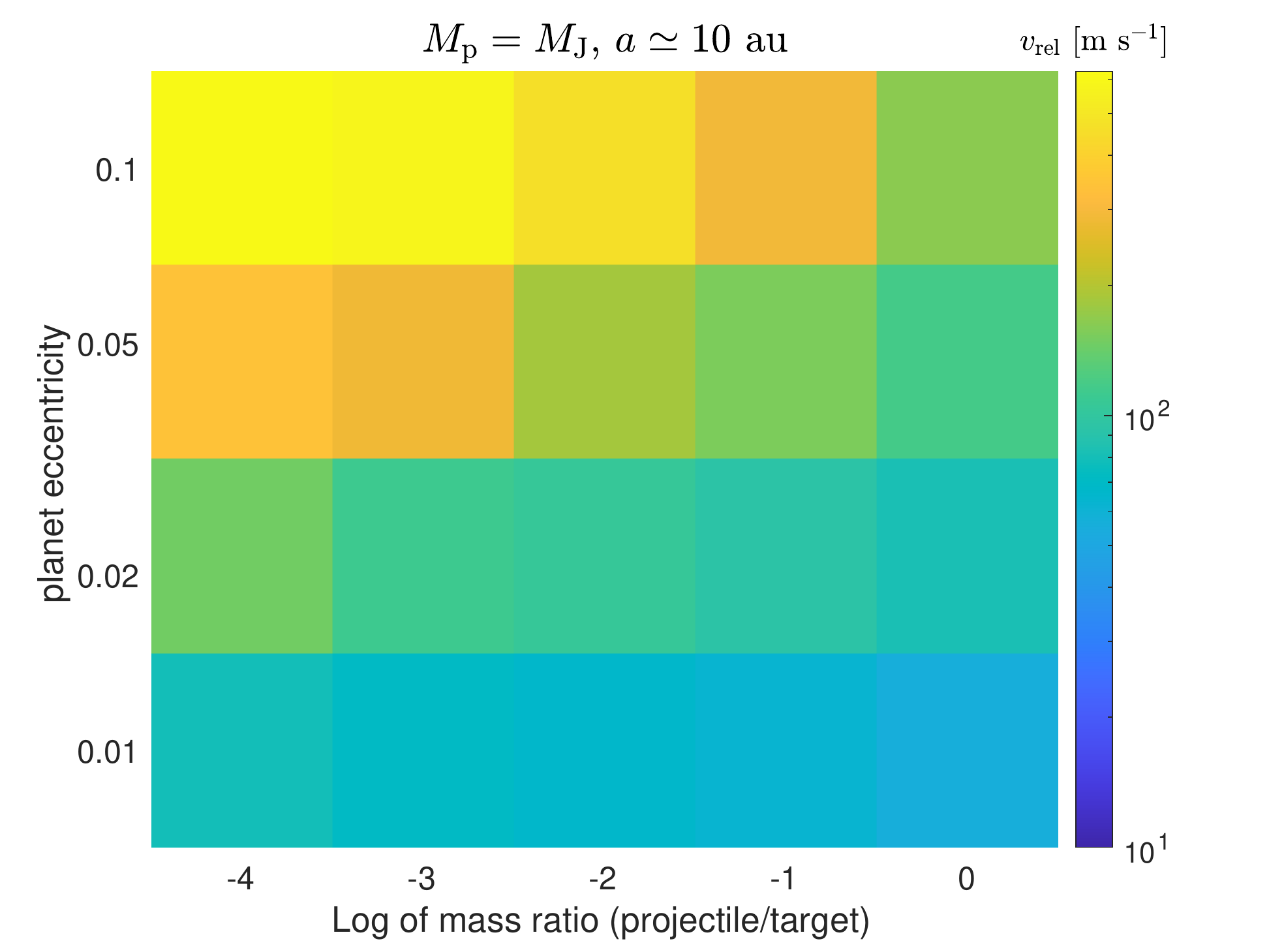}{0.45\textwidth}{(a)}
             {\fig{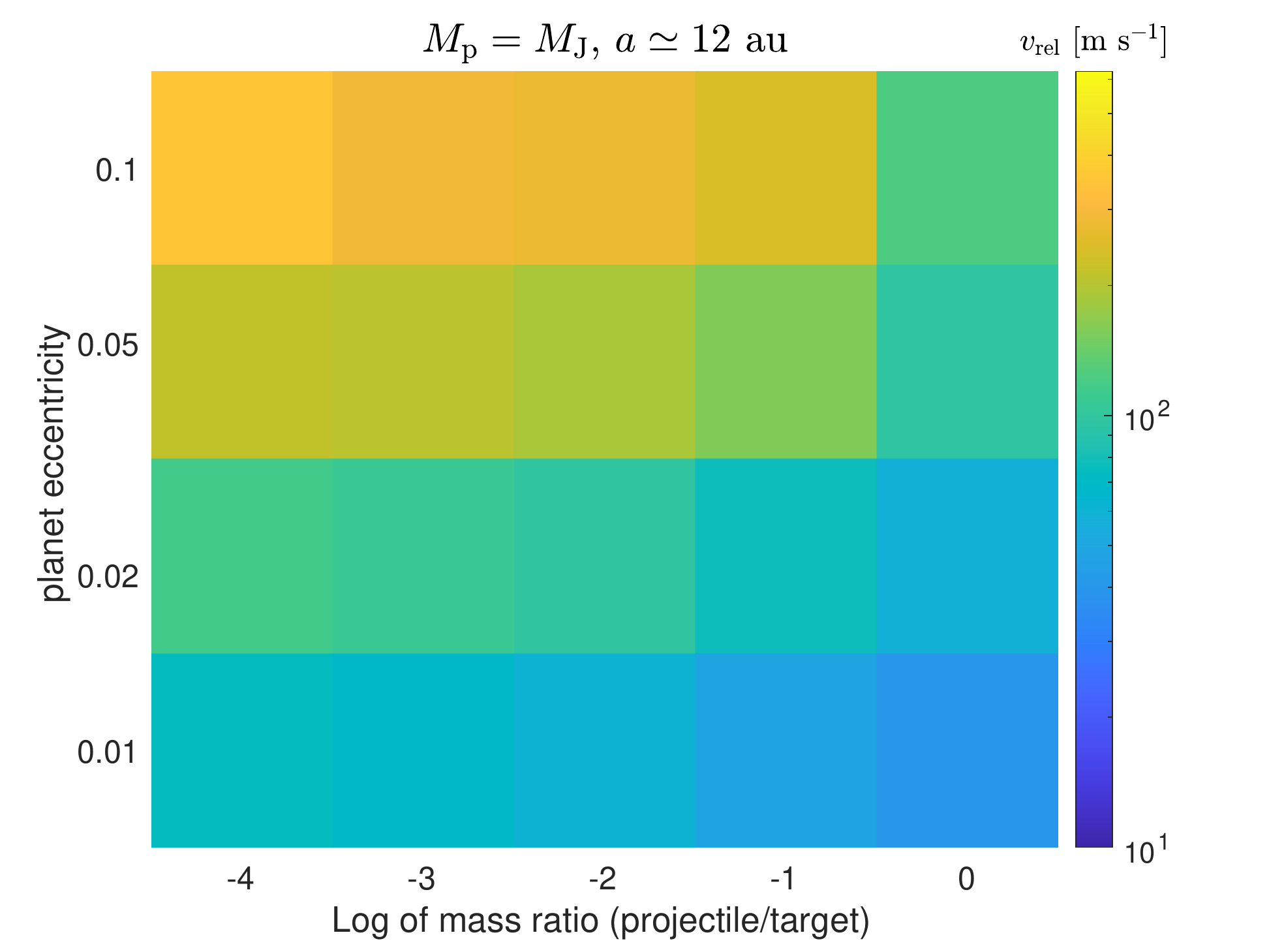}{0.45\textwidth}{(b)}}}
    \caption{Relative velocity between particles of different mass ratio for different planet eccentricity near (a) 10 and (b) 12 au. The color scale indicates the values of the relative velocities. The target mass is fixed as $m = 10^{20}~$g.}
    \label{fig:R-ep_out}
\end{figure*}

Fig. \ref{fig:R-ep_out} shows the dependence of the relative velocity between two populations of different masses on planet eccentricity. 
It is clear that $\langle \Delta v \rangle$ increases with increasing $e_{\rm{p}}$ for a given particle mass ratio.
Meanwhile, for a given planet eccentricity, $\langle \Delta v \rangle$ increases as the particle mass ratio increases.
Such trends remain the same for both values of the semi-major axis, with the $\langle \Delta v \rangle$ values generally being higher near 10 au.

\subsection{Growth limits and planetesimal sizes} \label{subsec:growth_limits}

\begin{figure*}
    \centering
    \includegraphics[width=0.9\textwidth]{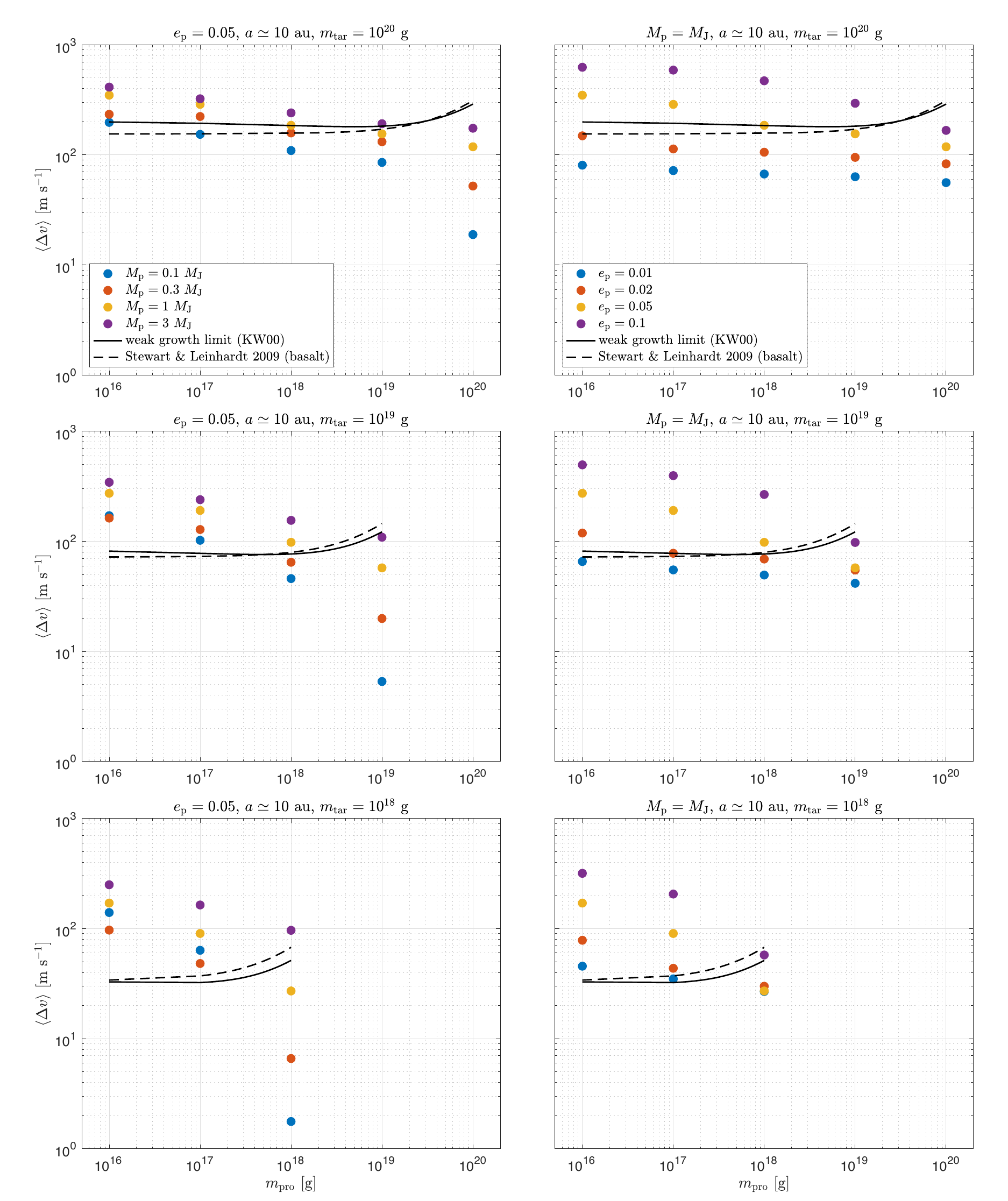}
    \caption{Relative velocities $\langle \Delta v \rangle$ between particles of given mass ratios near 10 au. Each row corresponds to a fixed target mass. The subplots in the left column shows the dependence of the relative velocities on the mass of the planet. Those in the right column shows the dependence of the relative velocities on the eccentricity of the planet. The solid curves represent the encounter velocity at the weak growth limit given in \citet{KW00} for reference. The dashed curves show the growth limit for basalt given in \citet{stewart2009velocity}.}
    \label{fig:dv_10au}
\end{figure*}

\begin{figure*}
    \centering
    \includegraphics[width=0.9\textwidth]{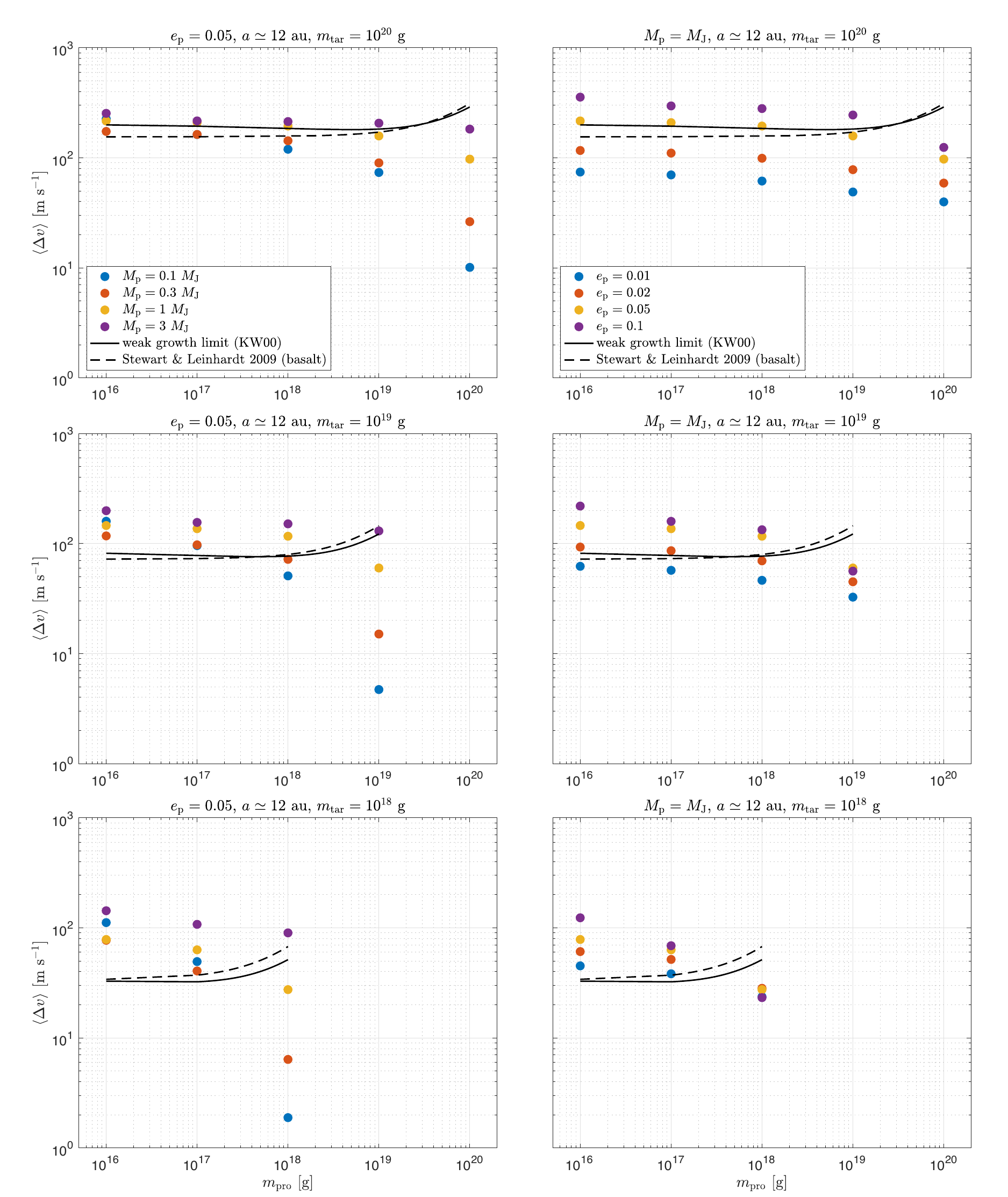}
    \caption{Same as Fig. \ref{fig:dv_10au} but for particles near 12 au.}
    \label{fig:dv_12au}
\end{figure*}

In this section, we discuss the dependence of the relative velocities on the size (mass) of planetesimals as well as other parameters (planet mass and eccentricity, semi-major axis), and compare the velocities with the growth limits, which mainly depend on the sizes of planetesimals. 

As we have mentioned in Section \ref{subsec:model}, we use a narrower mass range ($10^{16}$-$10^{20}~$g) for planetesimals compared with that in GK21 ($10^{13}$-$10^{20}~$g). 
The reason is that the dependence of planetesimal relative velocities and orbital alignment on the mass (ratio) of planetesimals have been elaborated in GK21, and in this paper we would like to stress the dependence of our results, in particular the planetesimal velocities, on the mass and eccentricity of the planet. 
Since the smaller particles ($m \lesssim 10^{15}~$g) drift inward due to gas drag on a rather short timescale (GK21), it is difficult to demonstrate the relative velocities between these particle populations and larger particles. 

Despite of the narrower mass spectrum, we can still infer from our results the dependence of the relative velocities on the mass (size) of planetesimals.
Fig. \ref{fig:dv_10au} and \ref{fig:dv_12au} summarizes the relative velocities of planetesimals with different mass ratios in all planet models. 
We compare the values of $\langle \Delta v\rangle$ with the growth limits in order to find out the favorable combinations of parameters for planetesimal accretion.
The growth limits are the critical encounter velocities calculated according to \citet{KW00} and \citet{stewart2009velocity}.
The criteria for the growth limits are when the ejected mass is equal to the projectile mass (solid line) and when the mass of the largest remnant is equal to the target mass (dashed line).

From the subplots in the left columns in Fig. \ref{fig:dv_10au} and \ref{fig:dv_12au}, we observe that as the planet mass increases, the dependence of $\langle \Delta v\rangle$ on the mass ratio between the projectile and target becomes weaker. 
This is consistent with the results shown in Fig. \ref{fig:R-Mp_out} as the smaller color gradient for larger planet masses.
For the largest planet mass ($M_{\rm{p}} = 3~M_{\rm{J}}$), the variations in $\langle \Delta v\rangle$ along the planetesimal mass ratio is very small (within an order of magnitude).
Such a weaker dependence can be easily explained by the distribution of the equilibrium eccentricity vector (see Fig. \ref{fig:eccvec_10au_Mp}). 
As the planet mass increases, the distribution of the equilibrium eccentricity vectors of different particle masses becomes more compact, resulting in a smaller $\Delta \boldsymbol{e}$, and thus a smaller $\langle \Delta v \rangle$.
In contrast, the dependence of $\langle \Delta v\rangle$ on the planetesimal mass ratio remains relatively flat as the planet eccentricity varies (the right columns in Fig. \ref{fig:dv_10au} and \ref{fig:dv_12au}).
For all four models with varying planet eccentricity, the $\langle \Delta v\rangle$ values decrease slightly as the target-projectile mass ratio decreases. 

Comparing the rows in Fig. \ref{fig:dv_10au} and \ref{fig:dv_12au}, we can identify the dependence on planetesimal size: when the target mass decreases, $\langle \Delta v\rangle$ exceeds the growth limits for a wider range of projectile mass as well as planet mass or eccentricity.
In other words, planetesimal accretion is possible for fewer parameter combinations (planetesimal mass ratio, planet mass / eccentricity) as the target mass decreases.
Therefore, generally speaking, larger planetesimal mass and smaller target-to-projectile mass ratio is more favorable for accretion.

The velocities of planetesimals near 10 au are in general slightly higher than the velocities of those near 12 au due to closer distance to the perturber and stronger gravitational perturbation (GK21).
However, for the purpose of identifying the combinations of parameters which are favorable for planetesimal accretion, such discrepancies are not very significant and are thus omitted for discussion here.

From Fig. \ref{fig:dv_10au} and \ref{fig:dv_12au}, we can see that the relative velocities of identical-size planetesimals are almost always lower than the growth limits except when $M_{\rm{p}} = 3~M_{\rm{J}}$ and $m \lesssim 10^{19}~$g.
However, planetesimals are expected to deviate from a perfectly uniform size distribution in reality.
A Jupiter-like planet ($M_{\rm{p}} = M_{\rm{J}}$, $e_{\rm{p}} = e_{\rm{J}}$) marginally allows planetesimals of mass $m \gtrsim 10^{19}~$g to grow.
For a planet eccentricity of $e_{\rm{p}} = 0.05$, to allow planetesimals with a size distribution $\log{(m_{\rm{tar}}/m_{\rm{pro}})} \geq 1$ to accumulate, the planet mass should be $\lesssim M_{\rm{J}}$ for target masses of $m_{\rm{tar}} \simeq 10^{20}~$g or $\lesssim 0.3~M_{\rm{J}}$ for target masses of $m_{\rm{tar}} \simeq 10^{19}~$g.
Similarly, when the perturbing planet is of Jupiter mass, the planet eccentricity should be $\lesssim 0.02$ and the planetesimals should be $\gtrsim 10^{19}~$g.
Planets larger or more eccentric than Jupiter would likely impede the accumulation of planetesimals with a realistic size distribution.



\section{Discussions} \label{sec:discussion}

\subsection{Implications on planetary systems}

Our results corroborate the speculations we made from the results of GK21, where we conjectured that the alignment of planetesimal orbits enforced by the coupling effect of Jupiter's secular perturbation and gas drag can aid the growth of Saturn's core, and that the effect of orbital alignment weakens so that the relative velocities of planetesimals increase as the planet mass increases. 
Because we are uncertain of the exact time at which Saturn's core begins to grow via planetesimal accretion, it is difficult to constrain the mass of Jupiter during the runaway growth phase of Saturn's core. 
In addition, since Jupiter's eccentricity is not constant over time - it oscillates periodically as a result of secular resonance with other planets in the Solar System \citep[e.g.,][]{Brouwer_vanWoerkom_1950, solar_system_dynamics}, we are also uncertain of Jupiter's eccentricity during the growth of Saturn's core. 
The mass and eccentricity of the young Jupiter could have both been smaller than the present-day values, so that at some sweet spot of semi-major axis, the planetesimal orbits were well-aligned and the relative velocities fell below the growth limit, and the accretion of Saturn's core could have been assisted.

Our results may elucidate on the architecture of exoplanetary systems. 
When the relative velocities of planetesimals are high, the growth of a planetary core is more likely to be impeded. 
Since both the increase in planet mass and eccentricity would result in higher relative velocity of planetesimals, we can infer that when a planet with a sufficiently large mass or eccentricity is present, we expect no subsequent planets to form in that system.
Recently, it has been reported that approximately 27\% of systems with cold giant planets have two such planets, and that some 11\% of the California Legacy Survey (CLS) Sun-like stars with one known cold giant should have an additional undetected cold giant planet companion \citep{Zhu_2022}.
Based on our results, we can anticipate the orbital features of these cold giant planets accompanied by one another: the inner planet should be of moderate mass and on an orbit of moderate eccentricity, and the outer planet should have formed sufficiently far from the inner planet such that the effects of secular perturbation remain low.



\subsection{Uncertainty of the growth limits and relative velocities}
\label{subsec:velocity_uncertainty}

The growth limits shown in Fig. \ref{fig:dv_10au} and \ref{fig:dv_12au} are based on previous studies that employ laboratory data for fitting the critical impact energy $Q^*$ required for catastrophic disruption of the target. 
A precise critical encounter velocity beyond which a collision results in no growth of the target, however, depends on multiple factors such as the material properties and the impact parameters, which are oversimplified in our models compared with the vast parameter space in reality. 

In addition to the uncertainty of the growth limits, the relative velocities we derive from numerical results are different from the actual encounter velocities of planetesimals on colliding orbits. 
Instead of tracking the encounter velocity between each colliding pair, which would require a large amount of computational resource, we statistically calculate the relative velocity of planetesimal swarms, which is determined by the distribution of their eccentricity vectors.
Such a calculation method provides good estimates of the average encounter velocity for a large population, but might result in slightly deviated results compared with the actual values. 
The number of particles in a given semi-major axis bin and the dispersion of their eccentricity vectors obviously depend on the bin width $\Delta a$, as mentioned in Section \ref{subsec:method_velocity}.
Varying the value of $\Delta a$ in a reasonable range does not change the qualitative dependence of planetesimal velocities on the planet parameters.
However, when compared with the growth limits, the velocities calculated with a statistical method might suffer from certain inaccuracy.
Therefore, we note that the quantitative comparison of our results with the growth limits should be treated with caution due to the uncertainties in both the growth limits and the planetesimal relative velocities calculated with a statistical approach.



\subsection{Planetesimal sizes}

We refer to GK21 for how we determined the planetesimal mass range in our models.
Briefly speaking, the reason we set the upper limit of the planetesimal mass to be $m = 10^{20}~$g is that the effect of planetesimal self-gravity might becomes significant. 
The timescale of viscous stirring is almost on the same order of magnitude as the gas damping timescale for the particle mass $m = 10^{20}~$g (see Fig. 3 in GK21).
Therefore, we claim that our assumption of test particles is only safe when $m \lesssim 10^{20}~$g. 

However, some recent studies have shown that planetesimals might have formed through the \textit{streaming instability},  which predicts that planetesimals are born large, with a typical size of $\sim 100~$km in diameter \citep[e.g.,][]{Youdin_2005,Johansen2007_nature,Johansen_2009}. 
Assuming a density of $\rho = 1~\rm{g}~cm^{-3}$, we consider a maximum planetesimal size of $\simeq 30~$km in radius in this paper.
The large initial size of planetesimals predicted by recent planetesimal formation models makes us curious about the dynamical behavior of these bodies in the scenario described in our paper. 
Our results show that although the velocity dispersion of larger planetesimals is higher due to less efficient gas damping (see Fig. \ref{fig:Mp_sigmaV_out} and Fig. \ref{fig:ep_sigmaV_out}), these larger bodies tend to be more favorable for accretion as they have relative velocities below the growth limits for a wider range of mass ratio (Section \ref{subsec:growth_limits}).
Based on these results, we can expect that planetesimals with masses $\gtrsim 10^{20}~$g can meet the condition for accretion for a wider range of parameters (i.e., larger planet mass or eccentricity, and larger target-to-projectile mass ratio).

One thing worth noticing is that the effect of self-gravity might become important for planetesimal masses $\gtrsim 10^{20}~$g.
As we have shown in GK21, the timescale of viscous stirring $\tau_{\rm{vs}}$ is one order of magnitude longer than that of gas damping and secular perturbation for particles of $m=10^{20}~$g, and $\tau_{\rm{vs}}$ keeps decreasing with increasing particle mass. 
Therefore, if larger planetesimals ($m \simeq 10^{21}$-$10^{22}~$g if assuming a density of $\rho = 1~\rm{g}~cm^{-3}$ for icy planetesimals) as predicted by the streaming instability are considered, the self-gravity of planetesimals should be taken into account (see Section \ref{subsec:limitations} for possible effects of planetesimal self-gravity).

\subsection{Limitations and future perspectives} \label{subsec:limitations}


As in GK21, a major limitation of this paper is that we neglect the mutual gravitational interaction between planetesimals. 
While the planetesimal self-gravity can heat up the disk by exciting the eccentricity of planetesimal swarms, it can also reinforce the alignment of orbits through gravitational viscosity (or two-body relaxation) \citep{tsukamoto2007formation}.
Such an effect is significant for massive planetesimals ($m \simeq 10^{23}$-$10^{24}~$g considered in \citet{tsukamoto2007formation}). 
For smaller planetesimals ($m \lesssim 10^{20}~$g), the effect of self-gravity is less prominent as the timescale of viscous stirring is much longer than that of other effects such as gas damping and secular perturbation. 
Therefore, it is interesting to investigate the dynamical behavior of planetesimals of intermediate masses ($m \simeq 10^{21}$-$10^{22}~$g) with the effect of self-gravity considered. 
A GPU-supported \textit{N}-body simulation code is currently under development.
We plan to explore the effect of planetesimal self-gravity by performing full \textit{N}-body simulations and follow the dynamical evolution of more massive planetesimals, which are favored by recent planetesimal formation models.

Another limitation lies in our assumption of an axisymmetric, circular gas disk. 
A gas disk can become eccentric if a giant planet is embedded in it, even if the planet is on a circular orbit. 
\citet{Hsieh_Gu_2014} found that with a planet of 5 Jupiter mass embedded in a gaseous disk on circular orbit at 5 au, the gas disk exterior to the planet becomes eccentric and precesses at a rate proportional to the longitude of ascending node of the planet.
In our model where the planet mass is large (particularly when $M_{\rm{p}} = 3\, M_{\rm{Jup}}$), the simplified gas disk profile might be inaccurate. 
Studies of protoplanetary disks in binary star systems have shown that asymmetries in the gas disk structure can significantly affect the velocities of planetesimals embedded within.
A precessing eccentric gas disk that deviates from circular streamlines might have less damping effect on the eccentricity of planetesimals, and thus result in weaker alignment of orbits and higher relative velocity of planetesimals in a circumprimary disk \citep{Paardekooper2008}.
For a circumbinary configuration, \citep{Marzari_et_al_2008} has shown that structures in the gas disk, such as spirals and density waves, also lead to less collimated pericenters of planetesimal orbits and higher relative velocities.

In addition, studies of planetesimal dynamics in binary star systems have pointed out that the self-gravity of the gas disk plays an important role in driving the velocities of planetesimals embedded within. 
\citet{Rafikov2013} shows that for the case of an axisymmetric circumstellar disk, the direct secular eccentricity excitation by the companion is strongly reduced by the rapid apsidal precession of planetesimal orbits induced by the disk gravity, so that the planetesimal velocities are lowered by an order of magnitude or more at 1 au.
Provided that the circumstellar disk has a small eccentricity and is massive ($\sim 0.1~M_{\odot}$), they show that such effect of the disk gravity eliminates the fragmentation barrier for in situ growth of small planetesimals ($\lesssim 10~$km) even at a wide separation of 2.6 au \citep{Rafikov2013}.
On the other hand, eccentric, asymmetric gas disks would generally lead to excitation of planetesimal eccentricities and high impact velocities, making in situ growth challenging in a perturbed environment \citep[e.g.,][]{Marzari2013,Lines2016}.
More recently, \citet{Silsbee2021} show that the apsidal alignment of a protoplanetary disk with the binary orbit is one of the critical conditions for successful planetesimal growth.
These studies demonstrate that the gravity of the gas disk can bring a significant impact on the dynamics of planetesimals.
Therefore, in our future research, we expect to consider more realistic gas disk models in order to better constrain the effect of the gas components.

\section{Conclusions}

Building on the work of GK21, we further explore the dependence of the relative velocity of planetesimals on the mass and eccentricity of the perturbing giant planet.
We vary the planet mass in the range $M_{\rm{p}} =$ [0.1, 0.3, 1, 3]$\,M_{\rm{J}}$ and the planet eccentricity in the range $e_{\rm{p}} = $ [0.01, 0.02, 0.05, 0.1] while maintaining the semi-major axis of the planet fixed at 5.2 au.
We found that, despite that the distribution of the forced eccentricities for planetesimals of different masses becomes more compact when the planet mass increases, the dispersion of each population of eccentricity vectors becomes larger as a result of stronger secular perturbation and short-term gravitational kick during each encounter with the planet.
Consequently, the relative velocity of planetesimals monotonically increases as the planet mass increases.
Such increasing trend is the same at disk locations near 10 and 12 au.

As the planet eccentricity increases, its influence on planetesimal relative velocity is more subtle compared with that of the increasing planet mass.
The dependence of the velocity dispersion of identical-mass particles on planet eccentricity is determined by the balance between the timescales of the secular perturbation and gas damping.
Generally speaking, for particles of smaller masses ($m \lesssim 10^{18}~\rm{g}$), the effect of gas damping is dominant, and the velocity dispersion barely depends on the planet eccentricity; for particles of larger mass ($m \gtrsim 10^{19}~\rm{g}$), the effect of secular perturbation is more prominent, and thus the velocity dispersion increases with increasing planet eccentricity.
The situation is different near 10 au and 12 au due to the existence of MMR, which would excite the random motion of particles, resulting in increased velocity dispersion.
The relative velocity between two populations of different masses slightly increases as the planet eccentricity increases, which results from the more scattered distribution of the equilibrium eccentricity vectors for different particle masses.
Compared with the increase caused by the increasing planet mass, the increase in the velocity dispersion of identical-mass particles caused by increasing planet eccentricity is less significant.
In other words, the relative velocity of identical-sized planetesimals has a stronger dependence on the planet mass.

When fixing the planet mass $M_{\rm{p}}$ (or eccentricity $e_{\rm{p}}$), we find that planetesimal accretion is challenging for a wider range of parameters [mass ratio, $e_{\rm{p}}$ (or $M_{\rm{p}}$)] as the target mass decreases. 
If a Jupiter-like planet is in presence in the disk, collisional growth is likely impeded for planetesimals smaller than $\simeq 10^{19}~$g.

\vspace{12pt}
\begin{center}
ACKNOWLEDGEMENTS
\end{center}

The numerical computations were conducted on the general-purpose PC clusters at the Center for Computational Astrophysics, National Astronomical Observatory of Japan. E.K. is supported by JSPS KAKENHI grant No. 18H05438.


\bibliography{main}{}

\begin{thebibliography}{}
\expandafter\ifx\csname natexlab\endcsname\relax\def\natexlab#1{#1}\fi
\providecommand{\url}[1]{\href{#1}{#1}}
\providecommand{\dodoi}[1]{doi:~\href{http://doi.org/#1}{\nolinkurl{#1}}}
\providecommand{\doeprint}[1]{\href{http://ascl.net/#1}{\nolinkurl{http://ascl.net/#1}}}
\providecommand{\doarXiv}[1]{\href{https://arxiv.org/abs/#1}{\nolinkurl{https://arxiv.org/abs/#1}}}

\bibitem[{{Adachi} {et~al.}(1976){Adachi}, {Hayashi}, \&
  {Nakazawa}}]{Adachi_et_al_1976}
{Adachi}, I., {Hayashi}, C., \& {Nakazawa}, K. 1976, Progress of Theoretical
  Physics, 56, 1756, \dodoi{10.1143/PTP.56.1756}

\bibitem[{Boley {et~al.}(2010)Boley, Hayfield, Mayer, \&
  Durisen}]{BOLEY2010509}
Boley, A.~C., Hayfield, T., Mayer, L., \& Durisen, R.~H. 2010, Icarus, 207,
  509, \dodoi{https://doi.org/10.1016/j.icarus.2010.01.015}

\bibitem[{{Boss}(1996)}]{Boss_1996}
{Boss}, A.~P. 1996, \apj, 469, 906, \dodoi{10.1086/177838}

\bibitem[{Boss(1997)}]{Boss_1997}
Boss, A.~P. 1997, Science, 276, 1836, \dodoi{10.1126/science.276.5320.1836}

\bibitem[{Boss(2011)}]{Boss_2011}
---. 2011, The Astrophysical Journal, 731, 74,
  \dodoi{10.1088/0004-637x/731/1/74}

\bibitem[{{Brouwer} \& {van Woerkom}(1950)}]{Brouwer_vanWoerkom_1950}
{Brouwer}, D., \& {van Woerkom}, A. J.~J. 1950, Astronomical papers prepared
  for the use of the American ephemeris and nautical almanac, 13, 81

\bibitem[{Charnoz \& Brahic(2001)}]{Charnoz2001}
Charnoz, S., \& Brahic, A. 2001, A\&A, 375, L31,
  \dodoi{10.1051/0004-6361:20010917}

\bibitem[{Charnoz {et~al.}(2001)Charnoz, Thébault, \&
  Brahic}]{Charnoz_et_al_2001}
Charnoz, S., Thébault, P., \& Brahic, A. 2001, Astronomy and Astrophysics,
  373, 683, \dodoi{10.1051/0004-6361:20010517}

\bibitem[{{Fletcher} {et~al.}(2019){Fletcher}, {Nayakshin}, {Stamatellos},
  {Dehnen}, {Meru}, {Mayer}, {Deng}, \& {Rice}}]{Fletcher_2019}
{Fletcher}, M., {Nayakshin}, S., {Stamatellos}, D., {et~al.} 2019, \mnras, 486,
  4398, \dodoi{10.1093/mnras/stz1123}

\bibitem[{{Guo} \& {Kokubo}(2021)}]{Guo_Kokubo_2021}
{Guo}, K., \& {Kokubo}, E. 2021, \aj, 162, 115,
  \dodoi{10.3847/1538-3881/ac0895}

\bibitem[{Hayashi(1981)}]{Hayashi_1981}
Hayashi, C. 1981, Progress of Theoretical Physics Supplement, 70, 35,
  \dodoi{10.1143/PTPS.70.35}

\bibitem[{{Hayashi} {et~al.}(1985){Hayashi}, {Nakazawa}, \&
  {Nakagawa}}]{Hayashi_1985}
{Hayashi}, C., {Nakazawa}, K., \& {Nakagawa}, Y. 1985, in Protostars and
  Planets II, ed. D.~C. {Black} \& M.~S. {Matthews}, 1100--1153

\bibitem[{{Hsieh} \& {Gu}(2012)}]{Hsieh_Gu_2014}
{Hsieh}, H.-F., \& {Gu}, P.-G. 2012, \apj, 760, 119,
  \dodoi{10.1088/0004-637X/760/2/119}

\bibitem[{{Johansen} {et~al.}(2007){Johansen}, {Oishi}, {Mac Low}, {Klahr},
  {Henning}, \& {Youdin}}]{Johansen2007_nature}
{Johansen}, A., {Oishi}, J.~S., {Mac Low}, M.-M., {et~al.} 2007, \nat, 448,
  1022, \dodoi{10.1038/nature06086}

\bibitem[{Johansen {et~al.}(2009)Johansen, Youdin, \& Low}]{Johansen_2009}
Johansen, A., Youdin, A., \& Low, M.-M.~M. 2009, The Astrophysical Journal,
  704, L75, \dodoi{10.1088/0004-637x/704/2/l75}

\bibitem[{{Kokubo} \& {Ida}(2012)}]{kokubo2012dynamics}
{Kokubo}, E., \& {Ida}, S. 2012, Progress of Theoretical and Experimental
  Physics, 2012, 01A308, \dodoi{10.1093/ptep/pts032}

\bibitem[{{Kokubo} \& {Makino}(2004)}]{Kokubo_Makino_2004}
{Kokubo}, E., \& {Makino}, J. 2004, \pasj, 56, 861,
  \dodoi{10.1093/pasj/56.5.861}

\bibitem[{{Kortenkamp} \& {Wetherill}(2000)}]{KW00}
{Kortenkamp}, S.~J., \& {Wetherill}, G.~W. 2000, \icarus, 143, 60,
  \dodoi{10.1006/icar.1999.6241}

\bibitem[{{Kortenkamp} {et~al.}(2001){Kortenkamp}, {Wetherill}, \&
  {Inaba}}]{Kortenkamp_et_al_2001}
{Kortenkamp}, S.~J., {Wetherill}, G.~W., \& {Inaba}, S. 2001, Science, 293,
  1127, \dodoi{10.1126/science.1062391}

\bibitem[{Lines {et~al.}(2016)Lines, Leinhardt, Baruteau, Paardekooper, \&
  Carter}]{Lines2016}
Lines, S., Leinhardt, Z.~M., Baruteau, C., Paardekooper, S.-J., \& Carter,
  P.~J. 2016, A\&A, 590, A62, \dodoi{10.1051/0004-6361/201628349}

\bibitem[{Liu {et~al.}(2022)Liu, Raymond, \& Jacobson}]{Liu2022}
Liu, B., Raymond, S.~N., \& Jacobson, S.~A. 2022, Nature, 604, 643,
  \dodoi{10.1038/s41586-022-04535-1}

\bibitem[{{Marzari} \& {Scholl}(2000)}]{Marzari_Scholl_2000}
{Marzari}, F., \& {Scholl}, H. 2000, \apj, 543, 328, \dodoi{10.1086/317091}

\bibitem[{Marzari {et~al.}(1997)Marzari, Scholl, Tomasella, \&
  Vanzani}]{Marzari1997}
Marzari, F., Scholl, H., Tomasella, L., \& Vanzani, V. 1997, Planetary and
  Space Science, 45, 337, \dodoi{https://doi.org/10.1016/S0032-0633(96)00138-9}

\bibitem[{{Marzari} {et~al.}(2008){Marzari}, {Th{\'e}bault}, \&
  {Scholl}}]{Marzari_et_al_2008}
{Marzari}, F., {Th{\'e}bault}, P., \& {Scholl}, H. 2008, \apj, 681, 1599,
  \dodoi{10.1086/588423}

\bibitem[{Marzari {et~al.}(2013)Marzari, Thebault, Scholl, Picogna, \&
  Baruteau}]{Marzari2013}
Marzari, F., Thebault, P., Scholl, H., Picogna, G., \& Baruteau, C. 2013, A\&A,
  553, A71, \dodoi{10.1051/0004-6361/201220893}

\bibitem[{Meschiari(2012)}]{Meschiari_2012}
Meschiari, S. 2012, The Astrophysical Journal, 761, L7,
  \dodoi{10.1088/2041-8205/761/1/l7}

\bibitem[{Murray \& Dermott(2000)}]{solar_system_dynamics}
Murray, C.~D., \& Dermott, S.~F. 2000, Solar System Dynamics (Cambridge
  University Press), \dodoi{10.1017/CBO9781139174817}

\bibitem[{Paardekooper {et~al.}(2012)Paardekooper, Leinhardt, Thébault, \&
  Baruteau}]{Paardekooper2012}
Paardekooper, S.-J., Leinhardt, Z.~M., Thébault, P., \& Baruteau, C. 2012, The
  Astrophysical Journal, 754, L16, \dodoi{10.1088/2041-8205/754/1/l16}

\bibitem[{Paardekooper {et~al.}(2008)Paardekooper, Thébault, \&
  Mellema}]{Paardekooper2008}
Paardekooper, S.-J., Thébault, P., \& Mellema, G. 2008, Monthly Notices of the
  Royal Astronomical Society, 386, 973,
  \dodoi{10.1111/j.1365-2966.2008.13080.x}

\bibitem[{{Pollack} {et~al.}(1996){Pollack}, {Hubickyj}, {Bodenheimer},
  {Lissauer}, {Podolak}, \& {Greenzweig}}]{Pollack_et_al_1996}
{Pollack}, J.~B., {Hubickyj}, O., {Bodenheimer}, P., {et~al.} 1996, \icarus,
  124, 62, \dodoi{10.1006/icar.1996.0190}

\bibitem[{Rafikov(2013)}]{Rafikov2013}
Rafikov, R.~R. 2013, The Astrophysical Journal, 765, L8,
  \dodoi{10.1088/2041-8205/765/1/l8}

\bibitem[{{Scholl} {et~al.}(2007){Scholl}, {Marzari}, \&
  {Th{\'e}bault}}]{Scholl_et_al_2007}
{Scholl}, H., {Marzari}, F., \& {Th{\'e}bault}, P. 2007, \mnras, 380, 1119,
  \dodoi{10.1111/j.1365-2966.2007.12145.x}

\bibitem[{Silsbee \& Rafikov(2021)}]{Silsbee2021}
Silsbee, K., \& Rafikov, R.~R. 2021, A\&A, 652, A104,
  \dodoi{10.1051/0004-6361/202141139}

\bibitem[{Stewart \& Leinhardt(2009)}]{stewart2009velocity}
Stewart, S.~T., \& Leinhardt, Z.~M. 2009, The Astrophysical Journal, 691, L133

\bibitem[{Thebault(2011)}]{Thebault2011}
Thebault, P. 2011, Celestial Mechanics and Dynamical Astronomy, 111, 29,
  \dodoi{10.1007/s10569-011-9346-2}

\bibitem[{Th{\'e}bault {et~al.}(2006)Th{\'e}bault, Marzari, \&
  Scholl}]{THEBAULT2006193}
Th{\'e}bault, P., Marzari, F., \& Scholl, H. 2006, Icarus, 183, 193 ,
  \dodoi{https://doi.org/10.1016/j.icarus.2006.01.022}

\bibitem[{Thébault \& Brahic(1998)}]{Thebault_1998}
Thébault, P., \& Brahic, A. 1998, Planetary and Space Science, 47, 233,
  \dodoi{https://doi.org/10.1016/S0032-0633(98)00080-4}

\bibitem[{Thébault {et~al.}(2008)Thébault, Marzari, \& Scholl}]{Thebault2008}
Thébault, P., Marzari, F., \& Scholl, H. 2008, Monthly Notices of the Royal
  Astronomical Society, 388, 1528, \dodoi{10.1111/j.1365-2966.2008.13536.x}

\bibitem[{Thébault {et~al.}(2009)Thébault, Marzari, \& Scholl}]{Thebault2009}
---. 2009, Monthly Notices of the Royal Astronomical Society: Letters, 393,
  L21, \dodoi{10.1111/j.1745-3933.2008.00590.x}

\bibitem[{Tsukamoto \& Makino(2007)}]{tsukamoto2007formation}
Tsukamoto, Y., \& Makino, J. 2007, The Astrophysical Journal, 669, 1316

\bibitem[{Walsh {et~al.}(2011)Walsh, Morbidelli, Raymond, O'Brien, \&
  Mandell}]{Walsh2011}
Walsh, K.~J., Morbidelli, A., Raymond, S.~N., O'Brien, D.~P., \& Mandell, A.~M.
  2011, Nature, 475, 206, \dodoi{10.1038/nature10201}

\bibitem[{Xie {et~al.}(2009)Xie, Zhou, \& Ge}]{Xie_2009}
Xie, J.-W., Zhou, J.-L., \& Ge, J. 2009, The Astrophysical Journal, 708, 1566,
  \dodoi{10.1088/0004-637x/708/2/1566}

\bibitem[{Youdin \& Goodman(2005)}]{Youdin_2005}
Youdin, A.~N., \& Goodman, J. 2005, The Astrophysical Journal, 620, 459,
  \dodoi{10.1086/426895}

\bibitem[{{Zhu}(2022)}]{Zhu_2022}
{Zhu}, W. 2022, arXiv e-prints, arXiv:2201.03782.
\newblock \doarXiv{2201.03782}

\end{thebibliography}
\bibliographystyle{aasjournal}



\end{document}